\documentclass[%
 reprint,
superscriptaddress,
nofootinbib,
 amsmath,amssymb,
 aps,
]{revtex4-2}

\pdfoutput=1

\usepackage{graphicx}
\usepackage{dcolumn}
\usepackage{bm}
\usepackage[caption=false]{subfig}
\usepackage[dvipsnames]{xcolor}
\usepackage{soul}
\usepackage{multirow}
\usepackage{array}
\newcolumntype{C}{>{\centering\arraybackslash}m{2.2cm}}

\definecolor{equations}{HTML}{194f8a}
\definecolor{citations}{HTML}{1d8a34}
\usepackage{hyperref}
\hypersetup{backref=true,       
    pagebackref=true,               
    hyperindex=true,                
    colorlinks=true,  
    urlcolor= black,                
    linkcolor= equations,                
    citecolor=citations,
}

\usepackage{array}

\newcolumntype{C}[1]{>{\centering\arraybackslash}m{#1}}

\AtBeginDocument{\hypersetup{pdfborder={1 1 1}}}
\begin{document}

\title{Constraints on $f(R)$ and nDGP Modified Gravity Model Parameters with Cluster Abundances and Galaxy Clustering}%

\author{Rayne Liu}
\thanks{Equal contribution.}
\affiliation{Department of Astronomy, Cornell University, Ithaca, NY 14853, USA}
\author{Georgios Valogiannis}
\thanks{Equal contribution.}
\affiliation{Department of Physics, Harvard University, Cambridge, MA 02138, USA}
\author{Nicholas Battaglia}
\affiliation{Department of Astronomy, Cornell University, Ithaca, NY 14853, USA}
\author{Rachel Bean}
\affiliation{Department of Astronomy, Cornell University, Ithaca, NY 14853, USA}

\date{\today}

\begin{abstract}
We present forecasted cosmological constraints from combined measurements of galaxy cluster abundances from the Simons Observatory and galaxy clustering from a DESI-like experiment on two well-studied modified gravity models, the chameleon-screened Hu-Sawicki $f(R)$ model and the nDGP braneworld Vainshtein model. 

A Fisher analysis is conducted using $\sigma_8$ constraints derived from thermal Sunyaev-Zel'dovich (tSZ) selected galaxy clusters, as well as linear and quasi-linear redshift-space 2-point galaxy correlation functions. We find that the cluster abundances drive the constraints on the nDGP model while $f(R)$ constraints are led by galaxy clustering. The two tracers of the cosmological gravitational field are found to be complementary, and their combination significantly improves constraints on the $f(R)$ in particular in comparison to each individual tracer alone. 

For a model of $f(R)$ with a General Relativity (GR) fiducial case ($f_{R0} = 0$), we find a $2$-$\sigma$ upper limit of $f_{R0}\leq5.68\times 10^{-7}$. For the well-studied log-based fiducial parameter value in $f(R)$, $\text{log}_{10}(f_{R0})=-5$, paired with the parameter value $n=1$, we find combined $1$-$\sigma$ constraints of $\sigma(\text{log}_{10}(f_{R0}))=0.12$ and $\sigma(n)=0.36$. For the nDGP model with fiducial $n_{\text{nDGP}}=1$ we find $\sigma(n_{\text{nDGP}})=0.087$. Our results present the exciting potential to utilize upcoming galaxy and CMB survey data available in the near future to discern and/or constrain cosmic deviations from GR. 

\end{abstract}
\maketitle

\section{Introduction}
\label{sec:intro}

The $\Lambda$CDM model accredits the acceleration of cosmic expansion \cite{Perlmutter:1998np, Riess:2004nr,Eisenstein:2005su, Percival:2007yw, Percival_2010, Kazin_2014, spergel2013widefield, Ade:2013zuv, Ade:2015xua} to the negative pressure exerted by an unknown dark energy, either as a cosmological constant $\Lambda$ with a canonical equation of state $w = -1$, or as a variable scalar field known as quintessence \cite{1988NuPhB.302..668W, PhysRevD.37.3406, COPELAND_2006}. However, more direct evidence for of the underlying nature of dark energy remains absent. The
cosmological constant explanation suffers from stark incompatibility since the value inferred from astronomical observations is $\sim$120 orders of magnitude smaller than that predicted in particle physics; the quintessence field theories attempting to resolve the discrepancy face subsequent fine-tuning problems \cite{RevModPhys.61.1}. 

Modified gravity (MG) theories attempt to avoid this extra energy component by explaining the accelerating universe with altering the standard theory of gravity, namely Einstein's General Relativity (GR), in large scales \cite{Koyama_2016, Ishak_2018, Ferreira_2019}. While GR has been meticulously tested with astrophysics on smaller scales, such as the solar system tests \cite{Will_2006} and strong-gravity tests via gravitational waves \cite{Abbott_2017, Savchenko_2017}, MG can potentially be applicable to larger cosmic scales with relatively weak gravitational fields. Nevertheless, such remarkable tests of GR on small scales have already imposed stringent constraints \cite{Lombriser_2016, Lombriser_2017, Sakstein_2017, Ezquiaga_2017, Creminelli_2017, Baker_2017}, leaving a limited parameter space for most MG models. Two particularly well-studied MG models that survive are the Hu-Sawicki $f(R)$ model series \cite{Hu_2007}, which feature a Chameleon mechanism, and the normal-branch Dvali-Gabadadze-Porrati braneworld model (nDGP) \cite{Dvali_2000}, which introduces a fifth dimensional force (Vainshtein mechanism). They successfully evade the above small-scale tests, while also reproducing an expansion history indistinguishable from $\Lambda$CDM. Hence, constraints via other independent observational quantities, especially the growth of the cosmic large-scale structure (LSS), are of crucial importance \cite{1309.5385}. Complementary to constraints via geometric distance measurements of the expansion history, LSS growth are very sensitive to the phenomenology of the cosmological MG models of interest \citep{Linder_2005,1309.5385,Ishak:2019aay}. 

Current and future LSS surveys will measure the abundance of galaxy clusters, as well as the 3-dimensional positions and velocities of galaxy halos. Such measurements are powerful probes of the LSS growth and clustering, and subsequently the nature of gravity and dark energy. In this work, we explore the constraining power of cluster abundances from upcoming observations of the thermal Sunyaev-Zel'dovich (tSZ) effect by the Simons Observatory \cite{SO} and galaxy clustering from spectroscopic observations by the Dark Energy Spectroscopic Instrument (DESI) \citep{Levi:2013gra}\footnote{\url{https://www.desi.lbl.gov/}}, both when considered independently and combined with each other. 

Galaxy clusters have long been regarded as a promising set of observables to test MG, and their abundances represented as number counts, as well as mass profiles, both serve as powerful tools. Potential constraints on MG using a wide variety of signal types have been considered, including X-rays \cite{Cataneo:2014kaa,He:2015mva,Li:2015rva,Sakstein_2016}, the tSZ effect \cite{Mak_2012,He:2015mva, 1802.02165,2019PhRvD.100f3529C}, and weak lensing \cite{1708.07502,Barreira:2015fpa}.
In this work, the constraints from abundances of galaxy clusters over the large linear scales is inferred through constraints on measurements of $\sigma_8$, the mean amplitude of matter energy density fluctuations. We base these constraints on the forecasted weak-lensing and CMB-halo lensing calibrated tSZ galaxy cluster abundances in \cite{1708.07502, 2019PhRvD.100f3529C}. 

Meanwhile, mapping out the three-dimensional clustering of galaxies across the cosmic history offers another window into the underlying physical processes, including the gravity models, that shaped the LSS. Building upon the legacy of the recently completed analysis by the Extended Baryon Oscillation Spectroscopic Survey (eBOSS) \citep{2021MNRAS.500..736B, 2020MNRAS.499.5527T,2020MNRAS.tmp.3648D}, DESI is expected to constrain the properties of gravity and dark energy at unprecedented levels of accuracy \citep{Alam:2020jdv}, in combination also with the next generation of cosmological surveys, such as Euclid \citep{Laureijs:2011gra}, the Vera C. Rubin Observatory Legacy Survey of Space and Time (LSST) \citep{Abell:2009aa,Abate:2012za}, the Nancy Grace Roman Space Telescope \citep{spergel2013widefield} and SPHEREx \citep{Dore:2014cca}. A necessary requirement for the optimal interpretation of this upcoming wealth of observational data is the ability to reliably model the clustering statistics in the variety of competing scenarios, in our case the landscape of MG models. When galaxies are identified through spectroscopic measurements, in particular, one needs to take into account not only the non-linear growth of structure, but also the fact that galaxy peculiar velocities induce an observed anisotropy in the clustering pattern, the Redshift-Space Distortions (RSD) \citep{10.1093/mnras/227.1.1,1992ApJ...385L...5H,Hamilton:1997zq}. In this study we capture these effects closely following the recent work of \citep{Valogiannis:2019nfz}, that employed the Gaussian Streaming Model (GSM) approach \citep{1995ApJ...448..494F,doi:10.1111/j.1365-2966.2011.19379.x,Wang:2013hwa} with Lagrangian Perturbation Theory \citep{Zeldovich:1969sb,1989A&A...223....9B,Bouchet:1994xp,Hivon:1994qb,Taylor:1996ne,Matsubara:2007wj,Matsubara:2008wx,2013MNRAS.429.1674C,Matsubara:2015ipa} in the context of MG \citep{Aviles:2017aor,Aviles:2018saf,1901.03763}, in order to successfully model the multipoles of the anisotropic correlation function of halos in theories beyond GR.

We forecast individual and joint constraints on MG from these two probes using Fisher analysis.  In addition to obtaining model-dependent constraints through a state-of-the-art treatment of the galaxy bias and RSD effects in MG, this work explores tests of gravity through the combination of these two complementary and promising probes of the LSS. 
Our paper is structured as follows: first, in Section \ref{sec:formalism}, we outline our theoretical and observational formalism and assumptions. We then present our analysis results in Section~\ref{sec:results}, with a subsequent discussion including implications for future work in Section~\ref{sec:disc}. The details of the particular model used to obtain the galaxy clustering covariance matrices are presented in Appendix \ref{App:AppendixA}.

\section{Formalism}\label{sec:formalism}


\subsection{Modified Gravity Models}
\label{sec:MGMod}
  
 We focus on two quintessential models in the literature of MG, the Hu-Sawicki $f(R)$ and the nDGP braneworld models, which respectively realize the Chameleon and Vainshtein classes of screening.

\subsubsection{$f(R)$ Hu-Sawicki model}
In the Hu-Sawicki $f(R)$ model, a non-linear modification function, $f(R)$, of the Ricci scalar is added to the standard Einstein-Hilbert action:
 
 \begin{equation}\label{MGaction}
    S = \int d^4x\sqrt{-g}\left [ \frac{R + f(R)}{16 \pi G}+\mathcal{L}_m \right ],
\end{equation}
 where $G$ is the Newtonian gravitational constant, $\mathcal{L}_m$ the matter Lagrangian, and $f(R)$ induces the accelerating universe instead of a cosmological constant $\Lambda$. Through a conformal transformation, (\ref{MGaction}) as the Einstein frame expression can be cast into the form of a scalar-tensor theory with the scalaron, $f_R \equiv \frac{df(R)}{dR}$, acting as the MG-induced degree of freedom \citep{Brax:2008hh}. Imposing an expansion history identical to $\Lambda$CDM in the high curvature limit, a present day value of the scalaron can be obtained as  \cite{Valogiannis:2019nfz}
 \begin{equation}\label{fR0}
    f_{R0} = -n\frac{c_1}{c_2^2}\left[\frac{\Omega_{m0}}{3(\Omega_{m0} + \Omega_{\Lambda 0})}\right]^{n+1},
\end{equation}
 where $\Omega_{m0}$ and $\Omega_{\Lambda 0}$ are the normalized density parameters for nonrelativistic mass and dark energy today; they, as $\Lambda$CDM parameters, appeared in (\ref{fR0}) as a result of our assumption on the  expansion history. Instead of using $c_1/c_2^2$ as the free parameter in (\ref{fR0}), the pair of $f_{R0}$ (typically $\left |f_{R0}\right|$ in the literature, and for the rest of this paper we only consider $f_{R0} \geq 0$) and $n$ are commonly used. We recover the $\Lambda$CDM (GR) model when $f_{R0} \rightarrow 0$, which is the case in regions of high Newtonian potential, where the chameleon field becomes very massive due to the effect of the screening mechanism \citep{PhysRevD.69.044026,PhysRevLett.93.171104}.
 
 Extensive studies of the Hu-Sawicki model in the past decade have tightened constraints  on the available parameter space of the model \cite{Burrage2018,2020arXiv200908743D}, but have still left the model observationally viable and theoretically attractive, as it is devoid of instabilities \cite{Clifton:2011jh}. For  these reasons, it serves as the ideal test bed for us to explore constraints on MG with upcoming surveys of the LSS and CMB.
 
 \subsubsection{nDGP model}
 
 The Dvali-Gabadadze-Porrati (DGP) model is a representative example of the Vainshtein screening mechanism \citep{VAINSHTEIN1972393,Babichev:2013usa}, and features a modification to gravity due to a large extra fifth dimension of spacetime. The modified Einstein-Hilbert action is in this case
 
 \begin{equation}
    S = \int d^4x \sqrt{-g}\left[\frac{R}{16\pi G} + \mathcal{L}_m\right] + \int d^5 x \sqrt{-g_5}\frac{R_5}{16\pi G r_c},
\end{equation}
where $R_5$ and $g_5$ denote respectively the corresponding Ricci scalar and metric determinant of the fifth dimension, and $r_c$ the cross-over distance, a characteristic scale below which GR model becomes four-dimensional. A more appealing self-accelerating DGP model branch (sDGP), which requires no dark energy, has been shown to suffer from undesirable instabilities \cite{0709.2399}, hence we study the ``normal" branch (nDGP) coupled with a dark energy component to match the desired $\Lambda$CDM expansion history, which still remains interesting due to prior simulation investments. In this case, the only
free parameter to constrain is $n = H_0r_c$ ($H_0$ being the Hubble constant), of which the extensively studied values are $1$ and $5$. GR is recovered when $n \rightarrow \infty$, corresponding to the presence of large gradients of gravitational forces in Vainshtein screening.

\subsection{Cluster Abundances and $\sigma_8$}
\label{sec:abundances}
The constraints by cluster abundances are modeled after results obtained from \cite{1708.07502}, in which a Fisher forecast based on tSZ-selected galaxy clusters from a CMB-S4-like experiment is extended to model-independent constraints on the time-evolution of $\sigma_8(z)$. Specifically, we use forecasted errors on $\sigma_8(z)$ from \citep{1708.07502}, which predicted tSZ cluster abundances for Simons Observatory and included mass calibrations from optical weak-lensing and CMB-halo lensing while marginalizing over $\Lambda$CDM and 7 cluster mass-observable scaling relation parameters \citep[for more details see][]{1708.07502}. The correlations between these mass-observable scaling relation and cosmological parameters are shown in Fig. 6 of \citep{1708.07502}. $\sigma_8$ is the amplitude of matter energy density fluctuations smoothed out over a scale of $8 Mpc/h$, and its evolution over redshift $z$ is a promising probe of structure growth in the linear density perturbation regime.

To predict $\sigma_8(z)$ from MG models, we calculate $\sigma_R$ through the standard deviation of the probability density function of the matter density fluctuations, convoluted with a spherical top-hat window function $W(\mathbf{r}, R)$ with radius $R$:
\begin{equation}
W(\mathbf{r}, R) = \frac{1}{4\pi R^3/3} = \begin{cases} 
          1, & \left | \mathbf{r} \right |\leq R, \\
          0, & \left | \mathbf{r} \right |>R.
       \end{cases}
\end{equation}
Fourier transforming, Parseval's theorem gives
\begin{equation}
{\sigma_R}^2(z) = \int_{0}^{\infty}\frac{P(k, z)}{2\pi^2}\left [ \frac{3j_1(kR)}{kR} \right ]^2k^2dk,
\label{s8_calc}
\end{equation}
where $P(k, z)$ is the matter power spectrum at wavenumber $k$ and redshift $z$, $j_1(kR)$ is the spherical Bessel function of the first kind, and $3j_1(kR)/(kR)$ is the Fourier transform of the window function. $8 Mpc/h$ is then assigned to $R$.

In general, the power spectrum in MG can be obtained from the $\Lambda$CDM one by considering the modifications to the linear growth factor $D(k, z)$:
\begin{equation}
    P_{\text{MG}}(k, z) =  P_{\Lambda\text{CDM}}(k,z = 0) \cdot \left (\frac{D_{\text{MG}}(k,z)}{D_{\Lambda\text{CDM}}(z = 0)} \right )^2.
\label{psgrowth}
\end{equation}
The growth factors, more commonly expressed as $D(k, a)$ ($a = 1/(1+z)$ being the scale-factor), are obtained by solving the modified linear density evolution equations, extracted from the work of \cite{1011.1257}:
\begin{equation}
\ddot{D} + 2H\dot{D} - 4\pi G \rho_m (1 + g_{\text{eff}})D = 0,
\label{growthdiffeq}
\end{equation}
where $H$ is the Hubble parameter, $\rho_m$ is the non-relativistic matter density, and dots are derivatives with respect to time $t$. The effective gravitational factor $g_{\text{eff}}$  for $f(R)$ \cite{Valogiannis_2017} is 
\begin{equation}
  g_{\text{eff}} = \frac{k^2}{3(k^2 + a^2m(a))^2}
  \label{eq:fRgeff}
  \end{equation}
with the associated mass term

\begin{eqnarray}
    m(a) &=& \frac{1}{c} \sqrt{\frac{\left[\Omega_{m0} + 4(1 - \Omega_{m0})\right]^{-(n+1)}}{(n + 1)|f_{R0}|}}\times \nonumber
    \\ 
    &&\sqrt{\left[\frac{\Omega_{m0}}{a^3} + 4(1-\Omega_{m0})\right]^{n + 2}},
    \label{eq:fRm}
\end{eqnarray}

and for nDGP,
\begin{equation}
    g_{\text{eff}} = \frac{1}{3\beta(a)}
    \label{eq:nDGPgeff}
\end{equation}
where
\begin{equation}
    \beta(a) = 1 + 2Hr_c \left ( 1 + \frac{\dot{H}}{3H^2}\right ) = 1 + 2\frac{H}{H_0}n \left ( 1 + \frac{\dot{H}}{3H^2}\right ).
    \label{eq:nDGPbeta}
\end{equation}

For more direct comparison with the standard $\Lambda$CDM model, we evaluate $\sigma_{8(\text{MG})}/\sigma_{8(\Lambda\text{CDM})}$. We obtain the $\Lambda$CDM linear matter power spectrum at $z=0$, $P_{\Lambda\text{CDM}}(k,z=0)$, from the Boltzmann code CAMB \cite{Lewis:1999bs,Lewis:2002ah,Howlett:2012mh} as a starting point, and then utilize (\ref{s8_calc}) and (\ref{psgrowth}) to determine $\sigma_{8(\text{MG})}/\sigma_{8(\Lambda\text{CDM})}$. Based on the assumptions in MG, the structure growth at early times should be indistinguishable from that in $\Lambda$CDM, hence we normalize the ratio to $1$ at redshift $z=10$, high enough to set the initial conditions of structure growth. We note that the growth factors between the GR limit solution of (\ref{growthdiffeq}) and the $\Lambda$CDM prediction from CAMB differ at the level far below the statistical uncertainities to affect the results of the Fisher analysis, and are corrected by normalization using the former.

Our solution of $D(k,z)$ (or $D(k, a)$) for the $f(R)$ model is checked against the work of \cite{1903.08798} in which the code for linear perturbation in MG is slightly modified for our purpose. For both the $\Lambda$CDM and the nDGP models, the growth factor $D(z)$ (or $D(a)$) are scale-independent, and are checked against the empirical fitting function proposed by \cite{Linder_2005}:
\begin{equation}
    g(a) \equiv D(a)/a = \text{exp}\left [ \int_{0}^{a}\frac{da'}{a'}\left [ {\Omega_m(a')}^\gamma - 1\right ]\right ],
\end{equation}
 where $\Omega_m(a) = \Omega_{m0} a^{-3}/(H/H_0)^2$, $\gamma$ is taken as 0.55 for $\Lambda$CDM, and 0.68 for both sDGP and nDGP with a modified expansion history \cite{Linder_2005}. This agreement remains stable when $\Omega_{m0}$ is varied in a small range around our fiducial value $\Omega_{m0} \sim 0.315$. In our work after the check, the $\Lambda$CDM expansion history (Hubble parameter) is imposed on the nDGP model.

Using the marginalized errors on $\sigma_8(z)$ to summarize the constraints from cluster abundances has the following advantages. There is not much information lost since we are examining the linear regime, despite the fact that the constraints are compressed into a root-mean-squared quantity as $\sigma_8$. Performing a Fisher analysis using $\sigma_8$ based on \cite{1708.07502} is not only faster, but also more conservative in the sense that it does not introduce extra degeneracy breaking as is the case for full Fisher analyses where $\Lambda$CDM and nuisance parameters are not already marginalized over.  

\subsection{Galaxy Clustering Correlations}
\label{sec:galclustering}

The LSS of the universe, as traced by the observed inhomogeneous clustering pattern of galaxies, has been formed by the non-linear gravitational collapse of the primordial density distribution. We can model the observed clustering statistics of galaxies in MG, by taking into account the crucial effects of clustering in the quasi-linear regime, large-scale galaxy bias and redshift space distortions (RSD). Our modeling procedure summarized below is tailored to DESI observations, and is heavily based upon the previous works of \citep{1901.03763,Valogiannis:2019nfz}.

In the intermediate, quasi-linear scales, higher order perturbation theory can substantially improve upon the accuracy of the simple linear treatment, allowing for a robust modeling of the clustering statistics, without the need to resort to computationally expensive N-body simulations. In this work we focus on the Lagrangian Perturbation Theory (LPT), in which the expansion parameter is a vector field, $\bold{\Psi}$, which displaces each fluid particle from its initial position, $\bold{q}$, to its final, Eulerian one, $\bold{x}(\bold{q},t)$, through the mapping:
\begin{equation}\label{Lagpos}
\bold{x}(\bold{q},t) = \bold{q} + \bold{\Psi}(\bold{q},t).
\end{equation}
The first order LPT solution is
the famous Zel'dovich approximation \citep{Zeldovich:1969sb}. In MG theories, an additional degree of freedom is present, altering the perturbed Einstein equations and the non-linear gravitational evolution of dark matter overdensities, and subsequently the framework of LPT, as detailed in \citep{Aviles:2017aor,Aviles:2018saf,Aviles:2018qot,1901.03763}.

The galaxies observed by surveys of the LSS do not perfectly trace the underlying dark matter density distribution, but rather are biased tracers of it \citep{1984ApJ...284L...9K}. In the simpler picture of linear perturbation theory, the large-scale overdensity of biased tracers (i.e. galaxies) is proportional to the underlying dark matter overdensity \citep{1988MNRAS.235..715E}, while a wide range of more sophisticated treatments have been developed in the literature \citep{Desjacques:2016bnm}. When working in Lagrangian space as is in this work, biased tracers are identified as regions of the primordial density field pre-selected by a biasing function, $F$, that depends on the local matter density \citep{Matsubara:2008wx,2013MNRAS.429.1674C}. Given the statistical nature of cosmic density fields, the simplest meaningful observable statistic (in the configuration space) is the two-point correlation function,  $\xi_X(r)$, of tracers correlated over a distance $r$:
\begin{equation}\label{xiX}
    \xi_X(r) := \langle \delta_X(\bold{x})\delta_X(\bold{x} + \bold{r})\rangle, 
\end{equation}
where the angle brackets denote an ensemble average. In Lagrangian space, the ``Convolution Lagrangian Perturbation Theory" (CLPT) \citep{2013MNRAS.429.1674C,Vlah:2015sea,Vlah:2016bcl} was shown to work particularly well at recovering the correlation function of halos from N-body simulations, in $\Lambda$CDM cosmologies. Building upon these works, \citep{Aviles:2018saf,1901.03763} then expanded CLPT in the case of MG theories and successfully recovered the real-space two-point correlation function of dark matter haloes across the parameter space of the $f(R)$ and nDGP MG scenarios.

In addition to imperfectly tracing the dark matter distribution of the cosmic web, galaxies identified through spectroscopic means are observed in redshift space, rather than in real space, which further distorts the observed clustering pattern, known as the Redshift Space Distortions (RSD) \citep{10.1093/mnras/227.1.1,1992ApJ...385L...5H,Hamilton:1997zq}. Due to its peculiar velocity about the Hubble flow, $\bold{v}(\bold{x})$, a galaxy with real space position $\bold{x}$ will be instead observed at a redshift space position:
\begin{equation}\label{rsdpos}
\bold{s} = \bold{x} + \frac{\hat{z}\cdot\bold{v}(\bold{x})}{a H(a)}\hat{z},
\end{equation}
with $H(a)$ the Hubble parameter evaluated at scale-factor $a$. As a consequence, the redshift-space 2-point correlation function for biased tracers
\begin{equation}\label{xiRSD}
\xi_X^s(\bold{r}) = \langle \delta_X(\bold{s})\delta_X(\bold{s+r})\rangle, 
\end{equation}
becomes directionally dependent, unlike the real-space expression (\ref{xiX}). In large linear scales, coherent infall leads to the ``Kaiser boost", an enhancement on the amplitude of the two-point correlation function, whereas in the non-linear scales, the random velocities within virialized structures lead to the "Fingers-Of-God" (FOG) suppression effect. 

The Gaussian Streaming Model (GSM) \citep{1995ApJ...448..494F,doi:10.1111/j.1365-2966.2011.19379.x,Wang:2013hwa} has been shown to be very successful in modeling the anisotropic RSD correlation function of halos, through a convolution of the halo real space correlation function with the probability velocity distribution of tracers, that is approximated as a Gaussian \citep{PhysRevD.70.083007}. In particular, given the real-space mean pairwise velocity along the pair separation vector of a pair of tracers, $v_{12}(r)$,
as well as its pairwise velocity dispersion, ${\sigma}^2_{12}(r)$, 
the GSM gives the expression for the anisotropic RSD correlation function:

\begin{eqnarray}\label{xiGSM}
1+\xi^s_X(s_{\perp},s_{\parallel}) &=&\int_{-\infty}^\infty \frac{d y}{\sqrt{2\pi\sigma_{12}^2(\bold{r})}} 
[1+\xi^r_X(r)] \times \nonumber
\\
&& \exp\left[-\frac{\left(s_\parallel-y-\mu v_{12}(r)\right)^2}{2\sigma^2_{12}(\bold{r})}   \right],
\end{eqnarray}
where $s_{\perp},s_{\parallel}$ are the perpendicular and parallel to the line-of-sight components of the redshift-space separation s, with $s=\sqrt{s_{\perp}^2+s_{\parallel}^2}$, $r=\sqrt{s_{\perp}^2+y^2}$ and $\mu=\hat{r}\cdot\hat{z}= \frac{y}{r}$. Using CLPT to model the 3 ingredients that enter the prescription (\ref{xiGSM}), $\xi^r_X(r),v_{12}(r),\sigma^2_{12}(\bold{r})$, and based on the MG implementations of \citep{Aviles:2018saf,1901.03763}, \citep{Valogiannis:2019nfz} was able to successfully model the simulated RSD correlation function of haloes in the $f(R)$ and nDGP gravity scenarios up to 1-loop order in PT. For the purposes of this work, we only use the first order LPT solution (Zel'dovich approximation \citep{Zeldovich:1969sb}) to evaluate the GSM ingredients, mainly because our model for the evaluation of the clustering covariance matrix does not incorporate non-Gaussian corrections, as we explain in Appendix~\ref{App:AppendixA}. 

In addition to the  LPT growth factors and the linear power spectrum, derived from the underlying cosmological model, we include two nuisance parameters in the modeling of the observed galaxy clustering correlation function using the CLPT $\&$ GSM framework laid out in this section.  

The first nuisance parameter sets the first-order Lagrangian bias parameter, $b_1^E(z)$, which we use to calculate the Lagrangian bias up to second order (for a 1-loop prediction). We assume a redshift evolution for the bias of the form $b_1^E(z) D(z)=$ constant, following \citep{Font-Ribera:2013rwa}, where $D(z)$, the linear growth factor under $\Lambda$CDM, is normalized to unity at $z=0$. In our analysis, we treat $b_1^E(z=0)$ as a nuisance parameter over which we marginalize separately for the two types of DESI-like galaxy samples, Luminous Red Galaxies (LRGs) and the Emission Line Galaxies (ELGs), with fiducial values of $b_1^E(z=0)=1.7$ and $0.84$ respectively. 

We do not treat the second order bias parameter, $b_2^E$, as a nuisance parameter in its own right, but rather determine it in terms of the $b_1^E(z)$ prediction through the fitting formula calibrated from N-body simulations \citep{Lazeyras:2015lgp},

\begin{equation}\label{biasfit}
b_2^E = 0.412 - 2.143 b_1^E + 0.929 (b_1^E)^2 + 0.008 (b_1^E)^3.
\end{equation}

Fixing $b_2$, rather than allowing it to vary independently, is motivated by previous work \citep{White:2014gfa}, in which the galaxy correlation function multipoles on the large scales on which we are focusing were found to be insensitive to the particular values of the higher order bias factors ($b_n \geq b_2$), which were fixed to the corresponding peak-background split (PBS) prediction. This fact was further confirmed in the context of MG models in \citep{Valogiannis:2019nfz}, with the only difference being that in the current work we use the improved empirical fit from Eqn. (\ref{biasfit}) to determine $b_2$, instead of the PBS prediction.

Finally, the Eulerian bias values $b_1^E$ and $b_2^E$ can be converted to their Lagrangian space equivalents through the known conversion relationships \citep{Mo:1995cs,Mo:1996cn}:
\begin{equation}\label{eq:biasL}
\begin{split}
    b_1^L &=  b_1^E -1, \\
    b_2^L &=  b_2^E - \frac{8}{21}b_1^L .
\end{split}
\end{equation}

Motivated by the findings in \citep{Valogiannis:2019nfz}, we only include the $b_1$ and $b_2$ bias terms (together with the FoG contribution) as these were shown to be sufficient to accurately capture the RSD correlation function from the N-body simulations for the MG models we consider. Further, since we focus on the linear and quasi-linear regimes, using the Zel'dovich approximation, we do not attempt to include the small-scale effects through the corresponding EFT corrections, as was done e.g. in \citep{Chen_2021} for GR and more recently \citep{Aviles:2020wme} for MG theories, but rather reserve these expansions for future work. We finally note that throughout this analysis the bias is taken to be scale-independent, a good approximation for our scales of interest, as was explained in \citep{Valogiannis:2019nfz} and also previously confirmed by simulations in \citep{Arnold:2018nmv}, for both GR and MG.

The second nuisance parameter we include is a constant offset, $\alpha_\sigma$, that needs to be added to the modeled galaxy pairwise velocity dispersion,
\begin{equation}\label{asEFT}
\sigma^2_{12} \rightarrow \sigma^2_{12} + \alpha_\sigma,
\end{equation}
such that the latter matches the observed prediction from simulations/observations at the large scale limit, as we found in \citep{Valogiannis:2019nfz}. This correction aims to capture unknown small-scale non-linear effects and is essentially the equivalent of the ``Fingers-of-God" free parameter, $\sigma_{\text{FoG}}^2$, that is commonly employed in the simple phenomenological dispersion models \citep{1983ApJ...267..465D,10.1093/mnras/258.3.581}. While, in principle, the free parameter $\alpha_\sigma$ is different for each of the two galaxy samples we consider, in \citep{Valogiannis:2019nfz} it is shown that  in the Zel'dovich regime, which we adopt in this work, the best-fit value of $\alpha_\sigma$ changes only by a very small amount across the range of MG models and halo mass ranges we examine. This motivates the use of a common fiducial value for both the ELG and the LRG galaxy samples, in all cases.

\begin{figure}
\centering
\includegraphics[width=0.45\textwidth]{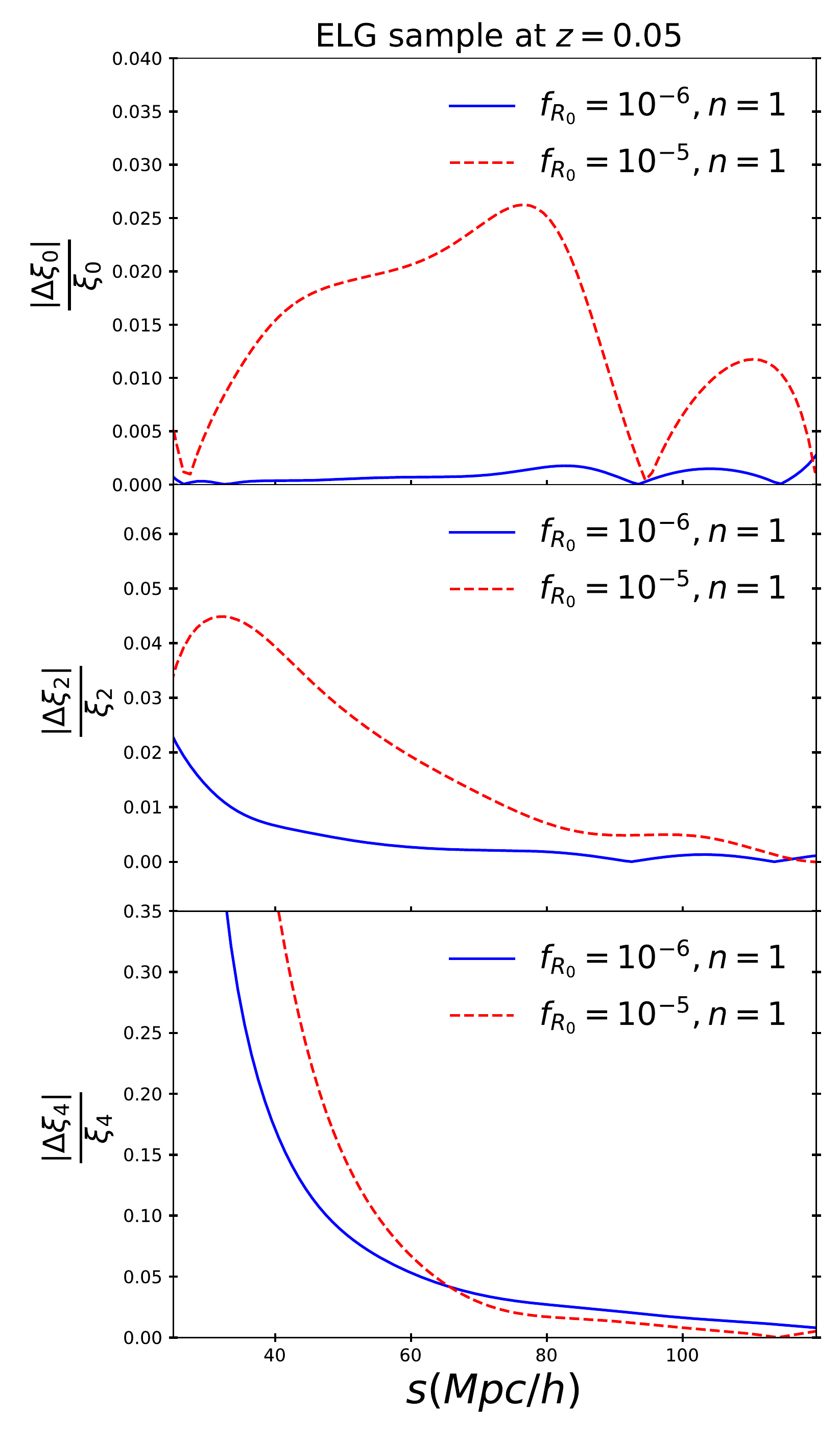}

\caption{The fractional deviations of the monopole [Top], quadrupole [Middle] and hexadecapole [Bottom] of the anisotropic correlation function in the $f_{R0}=10^{-5}$ [red dashed] and $f_{R0}=10^{-6}$ [solid blue] $f(R)$ scenarios w.r.t the corresponding $\Lambda$CDM prediction for the ELG sample at $z=0.05$.}
\label{fig:f(R)xicomp}
\end{figure}

All of those ingredients are combined to produce a prediction, by means of Eqn. (\ref{xiGSM}), for the MG RSD galaxy correlation function for the ELG and the LRG DESI galaxy samples at a given redshift $z$. As commonly performed in the literature, we further decompose the correlation function through a multipole expansion in a basis of Legendre polynomials, $L_l(\mu_s)$:
\begin{equation}
\xi(s,\mu_s)=\sum_l \xi_l(s) L_l(\mu_s),
\end{equation}
where the multipoles of order $l$ are given by
\begin{equation}\label{mult}
\xi_l(s)  = \frac{2 l +1}{2} \int_{-1}^{1} d\mu_s \xi(s,\mu_s) L_l(\mu_s),
\end{equation}
with $\mu_s = \hat{z}\cdot\hat{s} = s_{\parallel}/s$. We restrict our analysis on values $l=\{0,2,4\}$, which correspond to the monopole, the quadrupole and the hexadecapole, respectively (first 3 non-vanishing multipoles). In Fig.{~\ref{fig:f(R)xicomp}} we demonstrate an example of how the correlation function multipoles deviate from the corresponding $\Lambda$CDM prediction for the ELG sample at $z=0.05$, in the $f(R)$ scenario, for two sets of $f_{R0}$ and $n$ values. For a more thorough discussion of this topic, we refer the reader to \citep{Valogiannis:2019nfz, Aviles:2018saf, 1901.03763}.

\subsection{Fisher Analysis}
\label{sec:Fisher_method}

For the forecasted constraints on the two MG models we use Fisher analyses on the tracers of the LSS, e.g. cluster abundances in the linear regime and galaxy clustering in the quasi-linear regime. Our adopted fiducial background cosmology and MG parameters are shown in Table~\ref{tab:fidparams}, following \cite{1708.07502}. We consider three different $f(R)$ scenarios with fiducial values $f_{R0}=0$ (referred to as ``near-GR"), $10^{-5}$ (``F5"), and $10^{-6}$ (``F6"). In the near-GR case at $f_{R0} = 0$, the parameter $n$ is not well-defined and has no impact on the growth, so we consider its value as fixed at $n=1$ and do not include it as a Fisher parameter. Furthermore, we consider two nDGP scenarios with fiducial parameter values $n=\{1,5\}$, that we respectively refer to as N1 and N5.

To obtain constraints on $f_{R0}$, we use two different parameter space configurations. For the near-GR fiducial model, we use a linear parameter space, recognizing that we are probing a direct and comparatively small deviation from GR. Such a linear prior is used in the literature, see for example, ~Fig. 18 in \citep{2010.15278}, where $\omega^{-1}_{BD}$ is a generalized equivalent of $f_{R0}$. For F5 and F6 cases, on the other hand, we use a logarithmic prior, i.e. the corresponding parameter we use to constrain the models is $\text{log}_{10}(f_{R0})$. This is a common choice for constraining power-law-like parameters representing further deviations from GR, see \citep{1903.08798, Ramachandra:2020lue} for example.

In theory, the constraints from the two prior configurations are relatable through a direct transformation from the chain rule:
$\sigma[f_{R0}] = \sigma[\text{log}_{10}(f_{R0})] \cdot [f_{R0}\text{ln}(10)]$. Problems might arise on the lower limits of the constraints after such a transformation, which might reach down beyond the physically allowed range of $f_{R0} \geq 0$. Regarding this, we note that for each Fisher analysis, a Gaussian distribution is assumed of the likelihood around the fiducial value, i.e. a Gaussian distribution in the linear space around $f_{R0}$ for the near-GR case, and likewise in the log space around $\text{log}_{10}(f_{R0})$ for F5 and F6. A Gaussian distribution of constraints in the log space implies a skewed distribution of constraints linearly and vice versa. In order to keep our assumptions self-consistent, and supported by the literature as mentioned, we refrain from attempting a direct transformation of 1-$\sigma$ errors, and present our results in their respective prior spaces as-is. 

Another point of caution pertains to the case of fiducial models closer to GR, such as F6, for which the lower limit of the constraint can approach zero (as is the case with our result plots, Fig.s \ref{fig:F5_F6LRG+ELG_comp} and \ref{fig:constraint_plots_fR}). Our way of representing the results with both the upper and lower limits in the log prior, as well as the choice of the log prior itself in this case, are intended primarily to align with previous works in the literature as aforesaid  (e.g. see Fig. 10 of \citep{1903.08798}), in order to allow direct comparison with them. We note, however, that considering the attainable precision of experiments in the near future, it can be challenging to observationally distinguish the lower limit from zero, in which case the upper limit carries more observational value. 


\begin{table}[t!]
\begin{center}
\begin{tabular}{ |C{6em} | C{8em} |  C{6.0em} |}

\multicolumn{2}{c}{}
\\ \hline
 & \multirow{2}{*}{Parameter} & Fiducial  
\\
& &Value(s) 
\\ \hline
\multirow{5}{*}{$\Lambda$CDM} & $\Omega_ch^2$ & 0.1194 
\\ \cline{2-3}
& $\Omega_b h^2$ & 0.022 
\\ \cline{2-3}
& $H_0$ & 67.0 
\\ \cline{2-3}
& $10^9A_s$ & 2.2 
\\ \cline{2-3}
& $n_s$ & 0.96
\\\hline
\multirow{2}{*}{$f(R)$}& $f_{R0}$ & $0, 10^{-5}, 10^{-6}$ 
\\ \cline{2-3}
& $n$ & $1$ 
\\ \hline

nDGP & $n_{\text{nDGP}}$ & $1, 5$ 
\\ \hline
Nuisance & $b_1^E(z=0)$ & 1.7 (LRG) 0.84 (ELG)
\\ \cline{2-3}
 parameters & $\alpha_{\sigma}$ (in $Mpc/h$) & $0.5$  
\\ \hline
\end{tabular}
\caption{The fiducial cosmological parameters for the background cosmology considered in the analysis (denoted as $\Lambda$CDM) and the baseline $f(R)$ and nDGP MG scenarios. We add that the nuisance parameters refer only to the galaxy clustering evaluation. 
}
\label{tab:fidparams}
\end{center}
\end{table}

Under the Gaussian likelihood distribution assumption described previously, we employ the Fisher formalism \cite{0901.0721,0906.4123}:
\begin{equation}
    F_{ij} = \sum_{\alpha\beta}\frac{\partial f_\alpha}{\partial p_i}Cov^{-1}[f_\alpha,f_\beta]\frac{\partial f_\beta}{\partial p_j},
\label{Fisher}
\end{equation}
where $f_{\alpha, \beta}$ are the observables in bins labeled by $\alpha, \beta$; $Cov$ is the observable covariance matrix and $p_i$ and $p_j$ are a pair of the model parameters being constrained. 

The constraints by cluster abundances, as discussed in \ref{sec:abundances}, are represented by $\sigma_{8}(z)$. In particular, the observables $f_{\alpha}$ are the set of $\{\sigma_{8(\text{MG})}(z)/\sigma_{8(\Lambda \text{CDM})}(z)\}$ across $30$ linearly-spaced redshift bins centered on $z=0.05$ to $z=2.95$, which are predicted by the MG models. The observable covariance matrix $Cov^{-1}[f_\alpha,f_\beta]$ on $\sigma_{8}/\sigma_{8(\Lambda \text{CDM})}$, obtained from \cite{1708.07502}, is diagonal, and is a model-independent result where the errors from the background $\Lambda$CDM parameters are marginalized over, reducing the parameters to constrain in (\ref{Fisher}) to the pair $\{ f_{R0}, n\}$ ($f(R)$) or $n_\text{nDGP}$ (nDGP). Hence, the partial derivative stepsizes for the background cosmology parameters are aligned with that specified in \cite{1708.07502}. For the near-GR case, we take a stepsize of $10^{-8}$ directly with respect to $f_{R0}$, addressing a linear deviation from GR, with also noting that $\text{log}_{10}(f_{R0})$ is not well-defined. For F5 and F6, the partial derivatives are taken with respect to $\{\text{log}_{10}(f_{R0}),n\}$, with the respective stepsizes $\{0.05, 0.05\}$. In N1 and N5, the stepsize is $0.05$ for $n_\text{nDGP}$. A five-point central differences scheme is applied to evaluating the partial derivatives over all the MG parameters, giving a third-order accuracy, with the exception of the near-GR case where only a one-sided derivative is feasible due to the $f_{R0} \geq 0$ limitation. For this, a corresponding four-point forward differences scheme is then applied to maintain the third-order accuracy. The priors of the parameters for cluster abundances are inherited from \citep{1708.07502}, where information from the CMB is included and the priors are discussed and marginalized over (see around Table II of \citep{1708.07502}).

For galaxy clustering the observables are the galaxy correlation function multipoles, $f_{\alpha}=\{\xi_0(s),\xi_2(s),\xi_4(s)\}$, considered over $35$ spatial separation bins equally (logarithmically) spaced in the range $25<s<600 \, (Mpc/h)$. The cosmological parameters, $p_{i}$, are those given in Table~\ref{tab:fidparams}, while the covariance matrix for the monopole, quadrupole and hexadecapole moments is described in the Appendix~\ref{App:AppendixA}. Our evaluation assumes a DESI-like survey with LRG and ELG galaxy samples in the redshift range $0.15<z<1.85$, using $18$ linearly spaced $z$ bins, as outlined in \citep{Font-Ribera:2013rwa}. Our choices for the galaxy number density, survey volume and linear bias as a function of redshift are informed by \citep{Font-Ribera:2013rwa}, in particular its Table V. This assumes a mean galaxy number density of $\bar{n}_g \sim 2.5\times 10^{-4}$ $\left(Mpc/h \right)^{-3}$ and $\bar{n}_g \sim 5 \times 10^{-4}$ $\left(Mpc/h \right)^{-3}$, for LRGs and ELGs, respectively. The partial derivatives of the multipoles are evaluated with a  2-point central differences scheme, with the derivative step-sizes with respect to the background $\Lambda$CDM parameters and linear bias provided by~\cite{1708.07502}. 
With regards to the MG parameters, in the near-GR case with $f_{R0}\rightarrow 0$, we again employ a 3-point forward differences scheme, with a forward step of $10^{-8}$ in $f_{R0}$ while keeping $n$ fixed; the derivative steps for the $\{f_{R0},n\}$ pair in the F5 and F6 cases are $\{2\times10^{-6},0.4\}$ and $\{3\times10^{-7},0.4\}$, respectively, and for $n_{\text{nDGP}}$ in the N1 and N5 models they are $0.15$ and $0.5$. The step-size when differentiating with respect to the nuisance parameter $\alpha_{\sigma}$ is $1.5$, informed by the detailed study in \citep{Valogiannis:2019nfz}. We have checked that all of the choices above provide numerical stability in the derivatives. For galaxy clustering, no prior information was added for any of the parameters. The exploration of how to optimally include the BAO information, for example, would require additional appropriate priors (e.g. from Big Bang Nucleosynthesis) on the baryon density parameter $\Omega_b h^2$, a task we reserve for future study.

\section{Results}
\label{sec:results}

In this section we present the forecasted constraints on the cosmological parameters of Table~\ref{tab:fidparams} from galaxy clustering, and constraints on the MG parameters using combined observables, following the methods outlined previously.

\begin{figure*}
\centering
\includegraphics[width=0.92\textwidth]{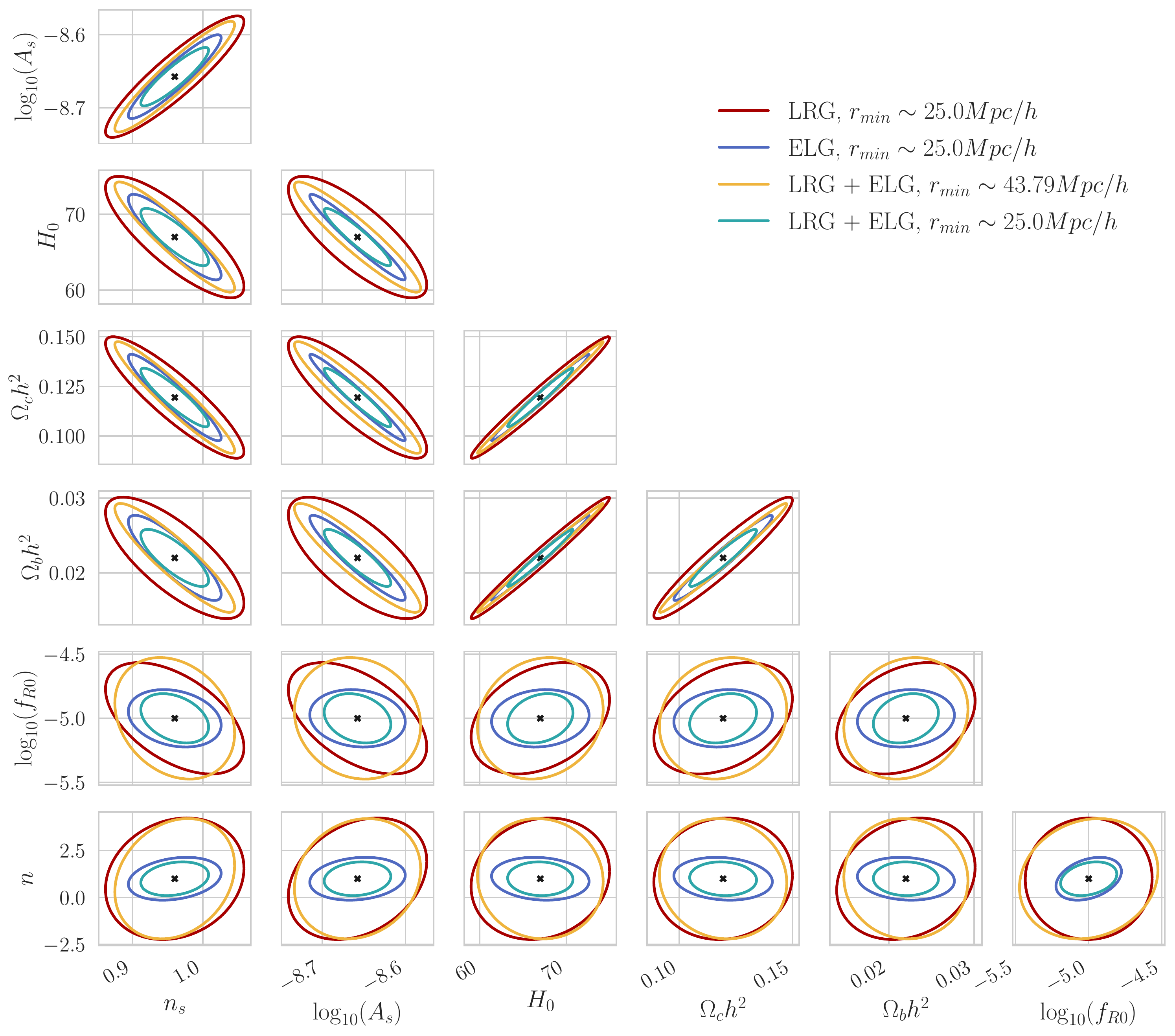}

\caption{Galaxy clustering constraints on the parameters in the F5 case (fiducial values $f_{R0} = 10^{-5}, n = 1$) using the DESI LRG sample [red], the ELG sample [blue] and both combined [cyan] for a minimum survey scale of $r_{\text{min}}\sim 25.0 Mpc/h$, the smallest scale we anticipate can be probed with the survey. Combined constraints for a more conservative minimum scale of $r_\text{min}\sim 43.79 Mpc/h$ [yellow] are also presented to show the impact of scale on the constraints.}
\label{fig:F5LRG+ELG_Full}
\end{figure*}

In Fig.~\ref{fig:F5LRG+ELG_Full}, we present 2-dimensional constraints from each parameter pair in the Fisher analysis, as obtained by the first three non-vanishing multipoles of the redshift-space correlation function of the LRG and ELG galaxy samples, both when considered separately and combined. In addition to the constraints on the standard $\Lambda$CDM parameters in line with previous works in the literature (e.g. \citep{Font-Ribera:2013rwa}), the complementarity of the LRG- and the ELG-derived contributions allows us to tightly constrain the pair of the MG parameters $\{\text{log}_{10}(f_{R0}),n \}$, which are the focus of our analysis. We see from the plot that the combined constraints from the two samples on the MG parameters are much tighter than the individual ones. It is also expected that using the ELGs produces tighter constraints in all parameter planes, given that this sample has a larger number density and redshift range, compared to the LRG counterpart (see Table V of \citep{Font-Ribera:2013rwa}). 


Fig.~\ref{fig:F5LRG+ELG_Full} also shows, as we further quantify in Table~\ref{tab:single_param_constraints_MG}, that the forecasted constraints on the MG parameter $\text{log}_{10}(f_{R0})$ are at least about an order of magnitude tighter compared to the parameter $n$. This finding is attributed to the fact that the 2-point function is more sensitive to variations of $f_{R0}$ than of $n$, in particular for the range of scales we consider in this work, as was found by the sensitivity analysis of \citep{Ramachandra:2020lue}. Due to this fact, most previous works in the literature (e.g. \citep{1903.08798,Bose:2020wch}) have commonly worked with a fixed value of $n=1$, and only considered constraints on $f_{R0}$. Thanks to our flexible analytical model for the anisotropic correlation function in any scalar-tensor theory, in this work we are able to provide constraints on the fuller parameter space of the $f(R)$ Hu-Sawicki model. 

In addition, in Fig.~\ref{fig:F5LRG+ELG_Full} we demonstrate how the choice of the minimum survey scale impacts the constraints we obtain, finding that a more conservative value of $r_\text{min}\sim 43.79 Mpc/h$ dilutes the constraining power overall. We find that the constraints from LRG+ELG combined for this more conservative value are comparable with the constraints from LRGs alone for $r_{\text{min}}\sim 25.0 Mpc/h$. This reduction in sensitivity is consistent with the predicted deviations in the Hu-Sawicki model becoming progressively more pronounced, as one considers  smaller scales. As a result, an analysis focusing on larger scales would restrict the ability to probe MG signals, resulting in looser constraints on the corresponding MG parameters. 

\begin{figure}
\centering
\includegraphics[width=0.39\textwidth]{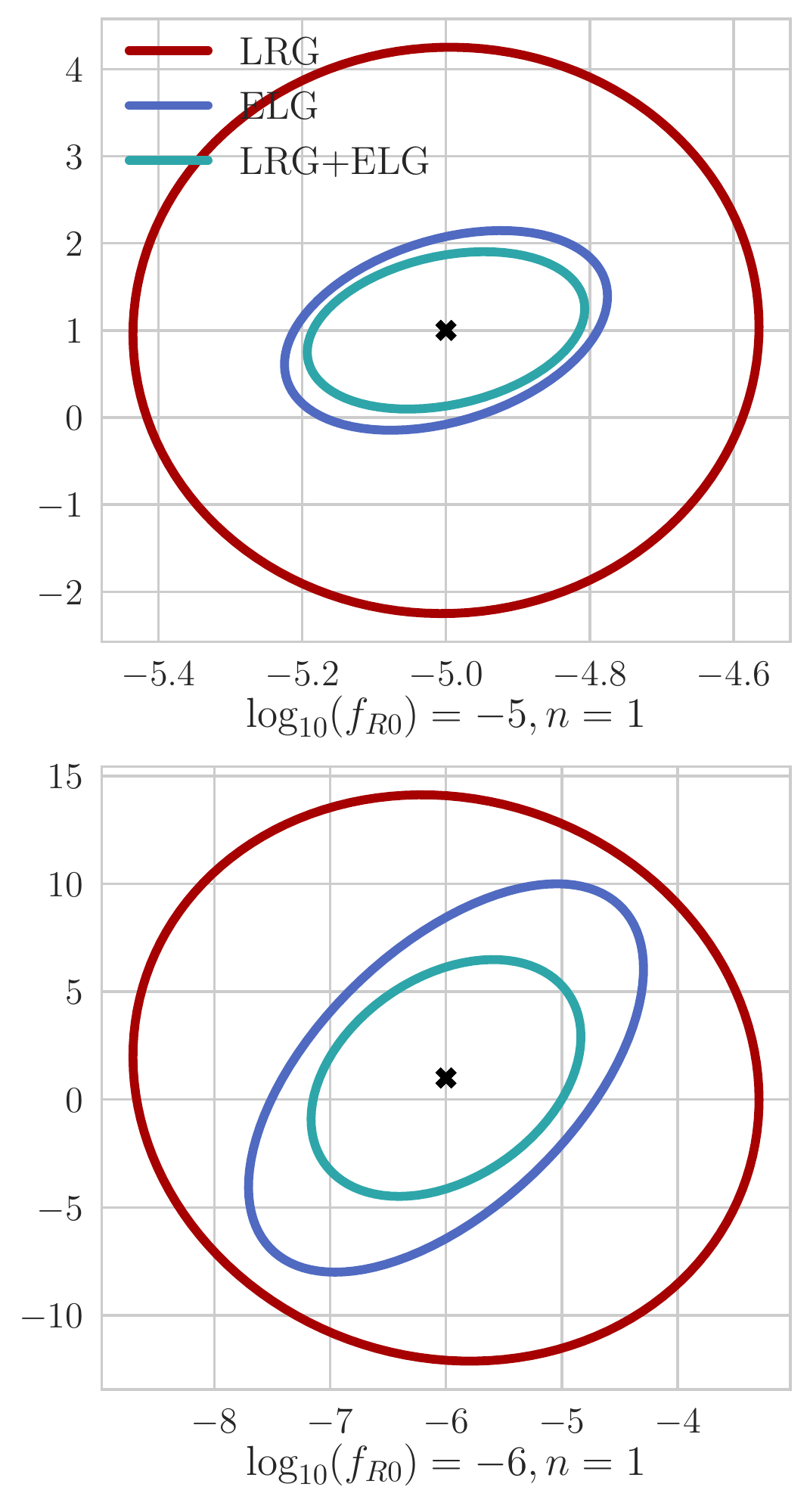}
\caption{Galaxy clustering constraints on $\{\text{log}_{10}(f_{R0}),n\}$ in the F5 (top) and F6 (bottom) cases (fiducial values $f_{R0} = 10^{-5}$ and $10^{-6}$, $n = 1$) for the LRG sample [red], ELG sample [blue] and both combined [cyan] assuming $r_{\text{min}}\sim 25.0 Mpc/h$. }
\label{fig:F5_F6LRG+ELG_comp}
\end{figure}

For the F6 model, we find that the 2D degeneracies between the $\Lambda$CDM and MG parameters are qualitatively similar to those for F5. In Fig.~\ref{fig:F5_F6LRG+ELG_comp} we show that the constraints on the two MG parameters are comparatively looser in F6 versus F5, consistent with the predicted deviations from GR being smaller. We find that the F6 constraints from LRG and ELG data are more comparable, with ELGs still being tighter, so that the combination of the two gives more notable relative improvements to the ELG data alone than for the F5 scenario. As we have discussed in section \ref{sec:Fisher_method}, for the F6 case the upper limit is more effectively represented under considerations of future experiments, and while we use the log space to ease comparison with the literature, we caution against the translation of the constraints to the linear space.

We also performed the same analysis on the nDGP models, but will only present the final combined constraints and omit showing the full corner plots for the sake of brevity. Our findings regarding nDGP are overall similar to the $f(R)$ scenario: the direction of the degeneracies for the background $\Lambda$CDM parameters are the same, while the degeneracies are stronger for the nDGP scenarios, presumably due to the scale-independence.

The 1D projected uncertainties obtained from galaxy clustering for the near-GR, F5, F6 $f(R)$ model and nDGP are summarized in Table~\ref{tab:single_param_constraints_MG}.  Across all the models considered, the uncertainties are principally driven by the ELG galaxy sample. For the near-GR case, the combination of LRGs with ELGs tightens the constraint on $f_{R0}$ from ELGs alone by 25\% (with the ELG constraint being about two-thirds that from LRGs). For F5, the ELG constraints on both parameters are roughly half the size of those from LRGs, and the combination of the two only reduces the uncertainties by 10$\sim$20\%. The impact of combining the two is more pronounced for F6, with ELG+LRG constraints about 30$\sim$40\% tighter than for ELGs alone. For both nDGP cases, the ELG constraint error on the $n_\text{nDGP}$ parameter is less than half of that from LRGs alone, and only a $\sim$10\% reduction is obtained by combining the two samples.

Our constraints for the $f(R)$ and nDGP cases are of the same order as the ones presented in \citep{Bose:2020wch}, which performed a Markov Chain Monte Carlo analysis.

\begin{figure*}
\centering
\subfloat[Varying $f_{R0}$, $n=1$]{%
  \includegraphics[scale = 0.31]{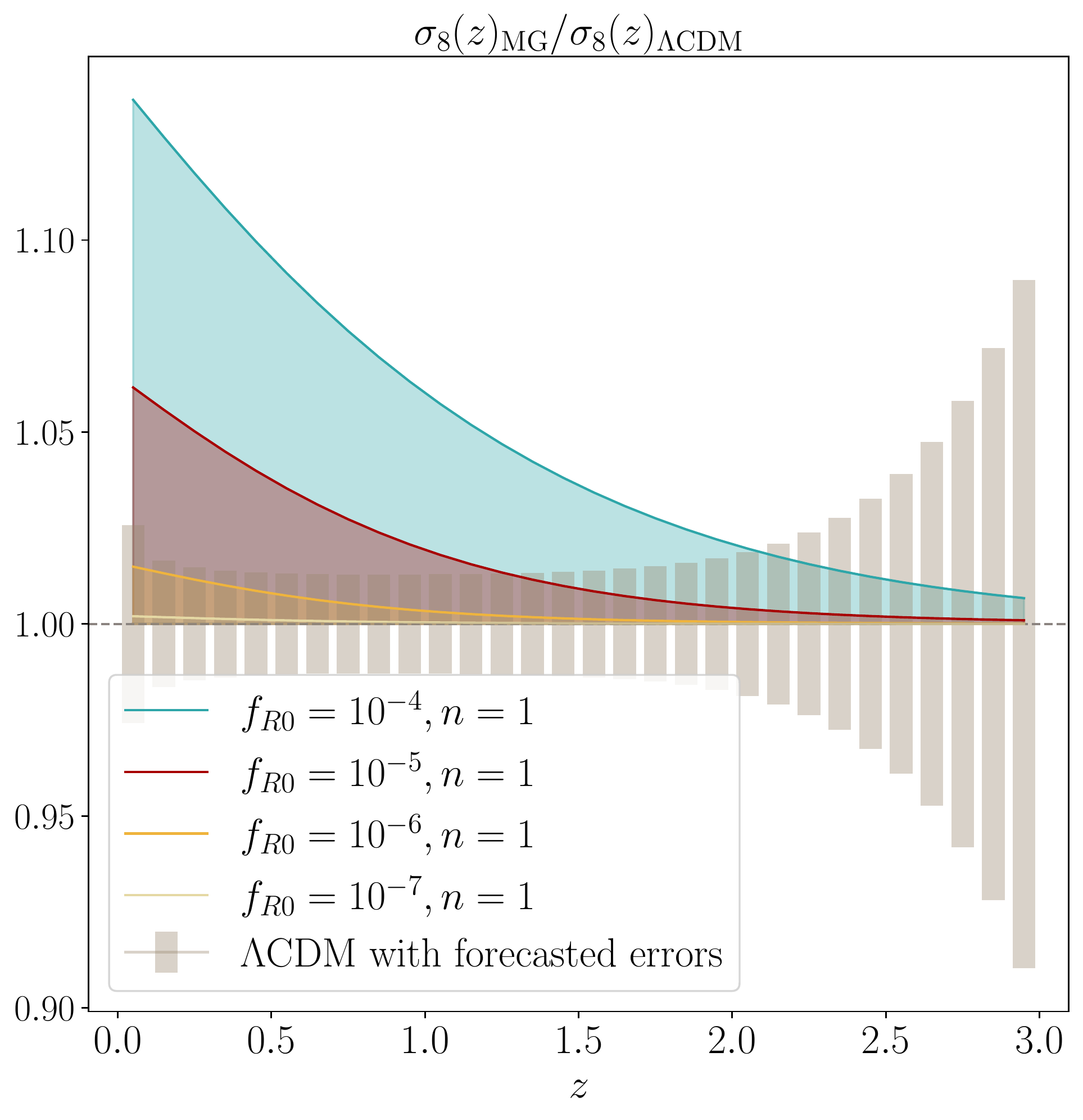}%
}
\subfloat[Varying $n$, $f_{R0} = 10^{-6}$]{%
  \includegraphics[scale = 0.31]{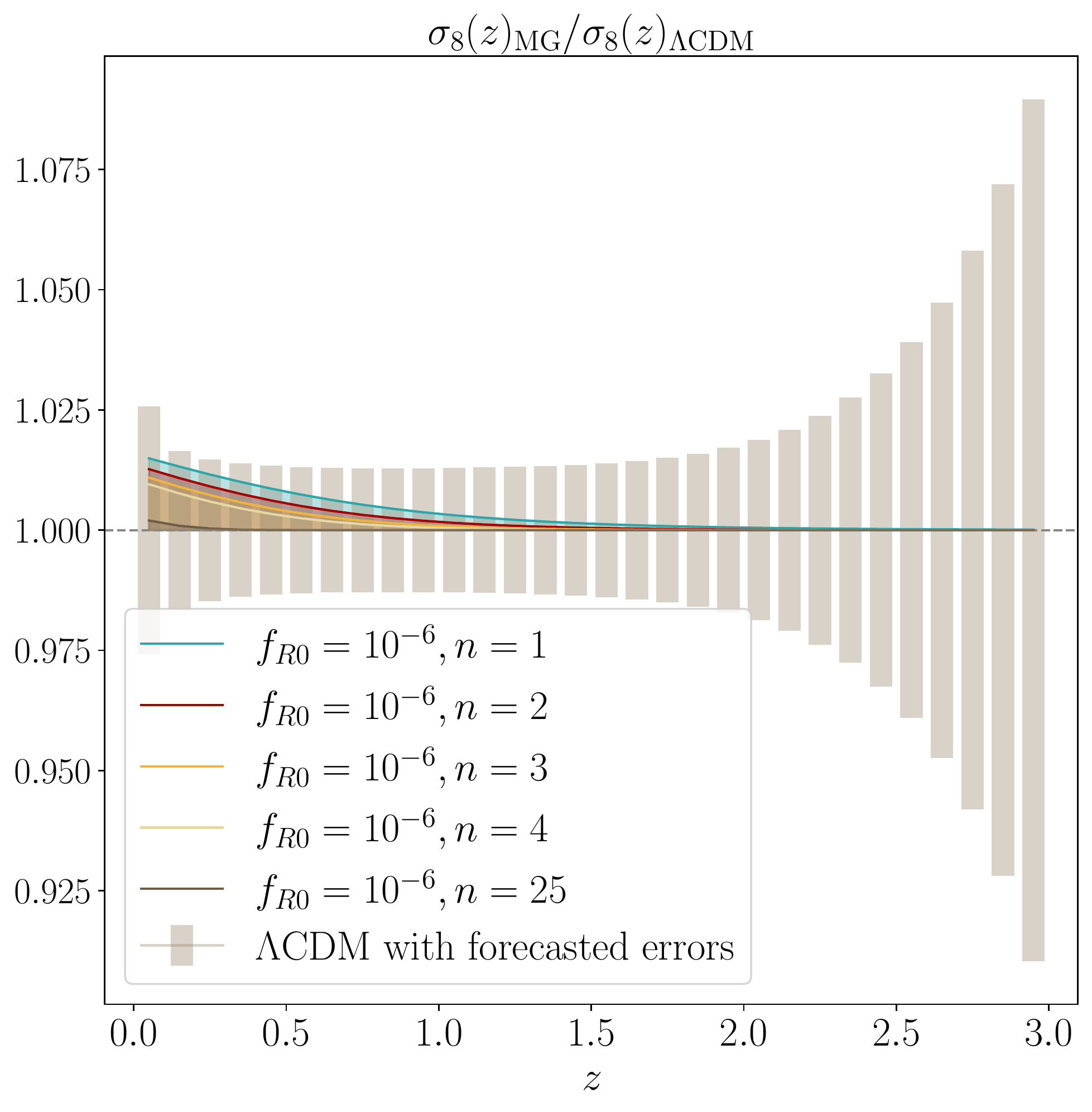}%
}
\subfloat[Varying $n$, for nDGP]{%
  \includegraphics[scale = 0.31]{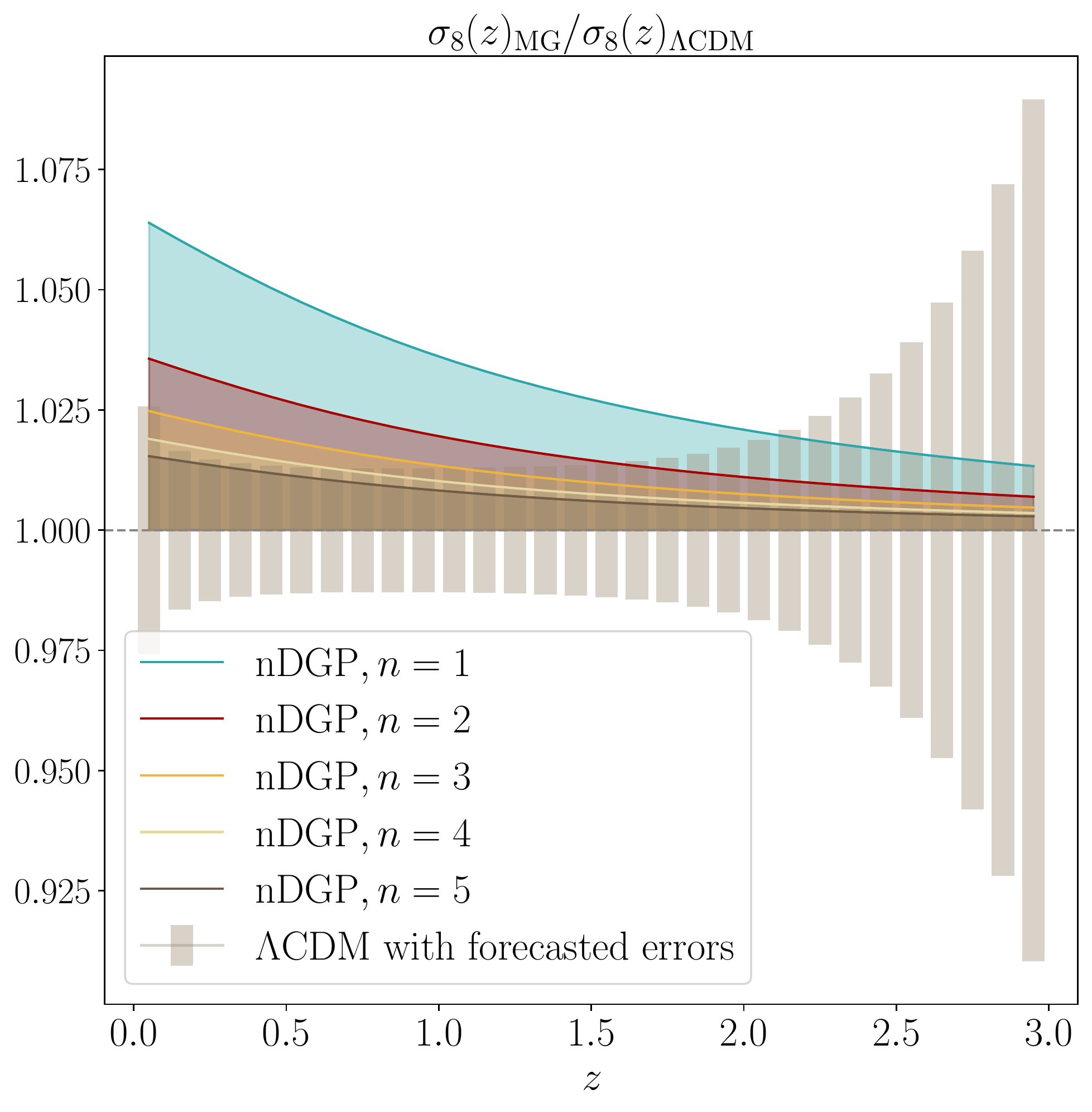}%
}

\caption{The MG predicted $\sigma_8(z)_{\text{MG}}/\sigma_8(z)_{\Lambda\text{CDM}}$ normalized at $z = 10$, plotted against the error of this ratio forecasted by cluster abundances. [Left, (a)] shows  scenarios with different $f_{R0}$ values while fixing $n$ in $f(R)$ as $1$. [Center, (b)] shows scenarios fixing $f_{R0} = 10^{-6}$ while changing $n$. [Right,(c)] shows the case in nDGP, where $n_{\text{nDGP}}$ is varied.}
\label{fig:MG overview}
\end{figure*}

We now consider the constraints from cluster abundances, obtained from $\sigma_{8(\text{MG})}/\sigma_{8(\Lambda\text{CDM})}$.  In Fig.~\ref{fig:MG overview}, we provide an overview comparison of the evolution of the predicted $\sigma_8$ ratio, over the 30 redshift bins from $z = 0.05$ to $z = 2.95$, for the MG models versus the  model-independent forecasted errors for future observations at the Simons Observatory \cite{SO}. The ratios predicted from MG are normalized at $z = 10$, consistent with the assumption that MG gives indistinguishable predictions from $\Lambda$CDM for LSS at high redshifts, varying $\{f_{R0}, n\}$ for $f(R)$ and $n$ for nDGP, respectively. The $\sigma_{8(\Lambda\text{CDM})}$ is normalized to be that calculated from (\ref{growthdiffeq}) (with $g_\text{eff} = 0$).

This overview provides an insight into the sensitivity of $\sigma_8$ with respect to the parameters of the two MG models we consider. For both the $f(R)$ and nDGP MG models, the constraining power mainly lies at lower redshifts, at $z<2$, increasing as one approaches $z=0$, where the deviation of the MG-predicted $\sigma_8$ is the highest, and the forecasted errors by cluster abundances are the tightest, notably for $0.5<z<1.5$. Furthermore, by comparing the signal to errors for the $f(R)$ case in sub-figures (a) and (b), we anticipate that the $\sigma_8$ data will be  more sensitive to variations in $f_{R0}$ than in $n$.

As is the case for galaxy clustering, we also summarize the model parameter constraints from galaxy cluster abundances in Table~\ref{tab:single_param_constraints_MG}. We find that for the near-GR case in $f(R)$, the constraints from cluster abundances and that from galaxy clustering are comparable. However, when it comes to constraining the model that deviates most greatly from GR, with a fiducial $\text{log}_{10}(f_{R0})=-5$, cluster abundances are not found to be as competitive as galaxy clustering. For this model, the constraint on $\text{log}_{10}(f_{R0})$ is a factor of 2.8 larger, while for $n$ it is 70\% larger. Cluster abundances then provide comparatively tighter constraints as we move to the weaker $f(R)$ model, with fiducial $\text{log}_{10}(f_{R0})=-6$, where the constraints on $\text{log}_{10}(f_{R0})$ and $n$, although effectively more of an upper limit, are roughly $10\%$ smaller than those predicted from galaxy clustering. 
We also find that cluster abundances provide constraints on the nDGP parameters that are roughly $50\sim60\%$ tighter than those from galaxy clustering.

\begin{table*}[t!]
\begin{center}

\begin{tabular}{|C{4em} |C{7em} | C{8em} ||  C{6.0em} |C{6.0em} |C{6.0em} ||C{6.0em}  ||C{6.0em} |}
\hline
\multicolumn{2}{|c|}{\multirow{2}{*}{Model}} & Fiducial 	&\multicolumn{3}{c||}{Galaxy clustering} &	Cluster	& \multirow{2}{*}{Combined}
\\ \cline{4-6}
\multicolumn{2}{|c|}{} &  Parameters						&	LRG  & ELG & ELG+LRG		 			& 	Abundances		&
\\ \hline \rule{0pt}{1.05\normalbaselineskip}
$2$-$\sigma$ & $f(R)$ near-GR &	$f_{R0}=0$ 				&   $\leq1.79\times 10^{-6}$	
&   $\leq1.15\times 10^{-6}$   	
&    	$\leq8.64\times 10^{-7}$
&	$\leq7.53\times 10^{-7}$
&  $\leq5.68\times 10^{-7}$
\\ \hline
\multirow{6}{*}{$1$-$\sigma$} & \multirow{2}{*}{$f(R)$ F5} &	$\text{log}_{10}(f_{R0})=-5$			&   $0.29$  	&  $0.15$	   	&    $0.13$ 				&	$0.37$				&  $0.12$
\\ 	
&  &	$n=1$							&   $2.14$  	&   $0.75$	   	&    	$0.59$ 				&	$1.00$				&  $0.36$
\\ \cline{2-8}
& \multirow{2}{*}{$f(R)$ F6} &	$\text{log}_{10}(f_{R0})=-6$			&   $1.78$ 	&   $1.12$	   	&    	$ 0.77$				&	$0.69$				&  $0.48$
\\
&  & $n=1$	& $8.64$	& $5.92$ 	&  $3.61$	&	$3.31$	& $2.30$
\\ \cline{2-8} \rule{0pt}{1.05\normalbaselineskip}
& nDGP N1& 	$n_{\text{nDGP}}=1$						&    $0.59$ 	&   $0.25$   	&    $0.23$	 					&		$0.094$			&  $0.087$
 \\ \cline{2-8} \rule{0pt}{1.05\normalbaselineskip}
& nDGP N5 &	$n_{\text{nDGP}}=5$						&   $8.30$  	&   $3.63$	   	&    	$3.29$ 					&	$1.77$				&  $1.56$
 \\ \hline
  	\end{tabular}

\caption{Marginalized one-parameter errors in MG models, presented using galaxy clustering (minimum survey scale $r_\text{min} \sim 25.0Mpc/h$) and cluster abundances alone respectively, and cross-combining the two observables. The numerical values within the same row of a fiducial parameter denotes the $1$-$\sigma$ ($68\%$ confidence) errors on the same parameter around that fiducial value. Specifically for the near-GR case (the top-row for the $f_{R0} = 0$ fiducial value), the 2-$\sigma$ ($95\%$ confidence) upper limits are reported.}
\label{tab:single_param_constraints_MG}
\end{center}
\end{table*}

\begin{figure*}
 \subfloat[$f_{R}$: near-GR (fiducial $f_{R0}=0$)]{%
 \includegraphics[width = 0.332\textwidth]{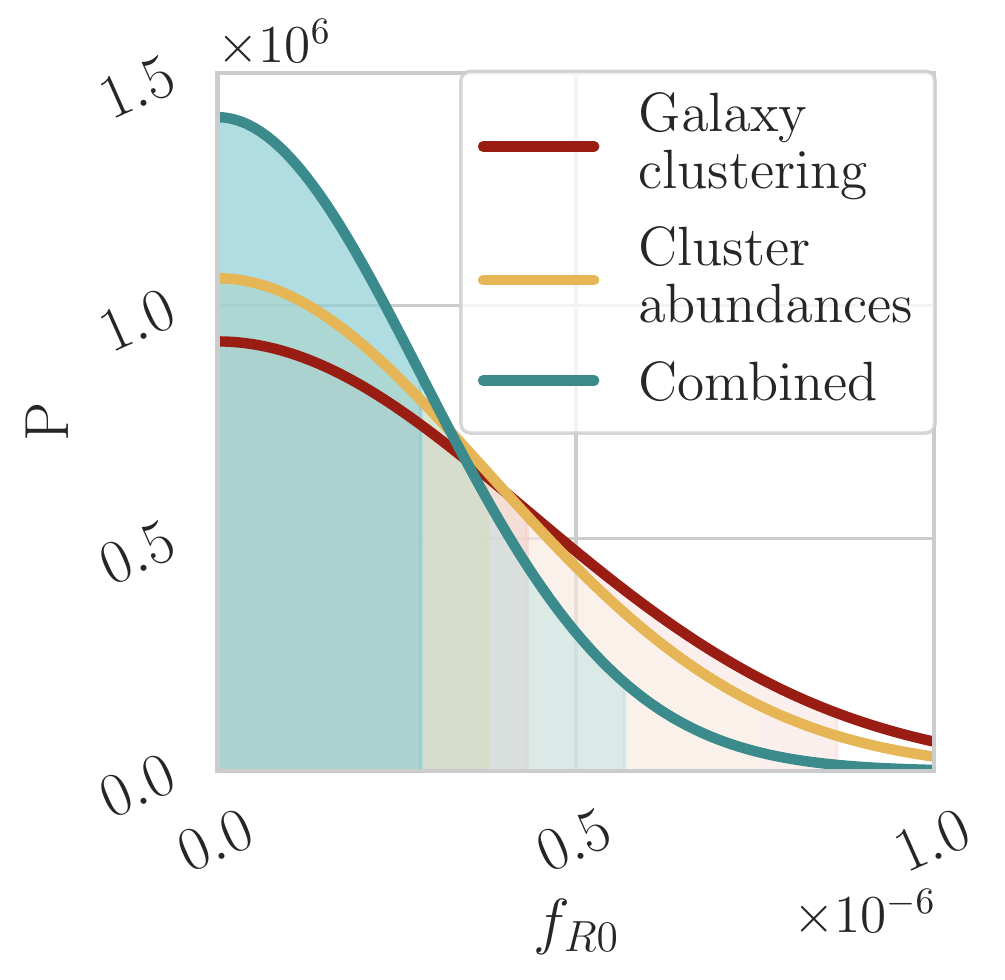}
 }
 \subfloat[nDGP: fiducial $n_{\text{nDGP}}=1$]{%
  \includegraphics[width = 0.3\textwidth]{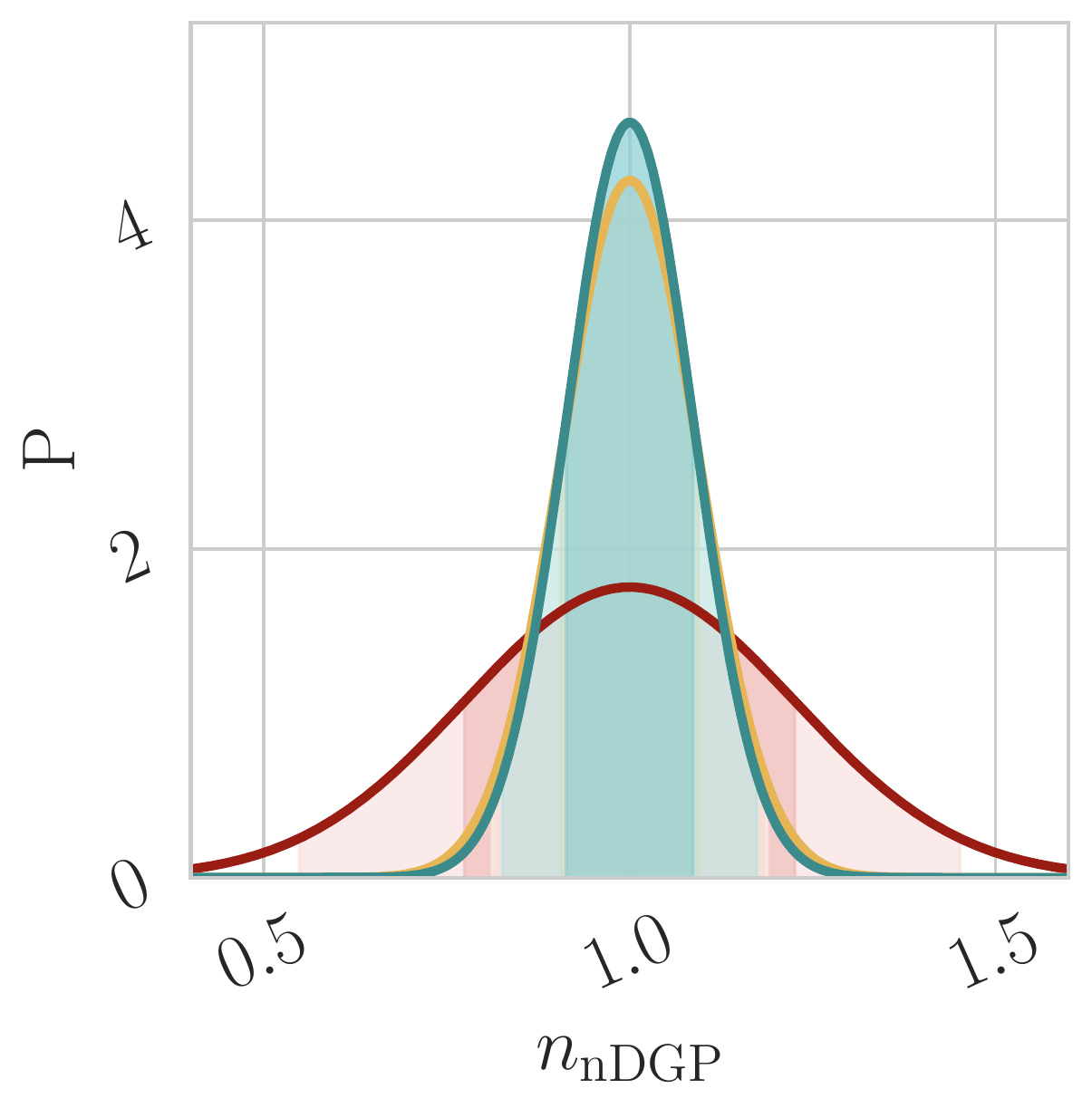}
  }
 \subfloat[nDGP: fiducial $n_{\text{nDGP}}=5$]{%
 \includegraphics[width = 0.313\textwidth]{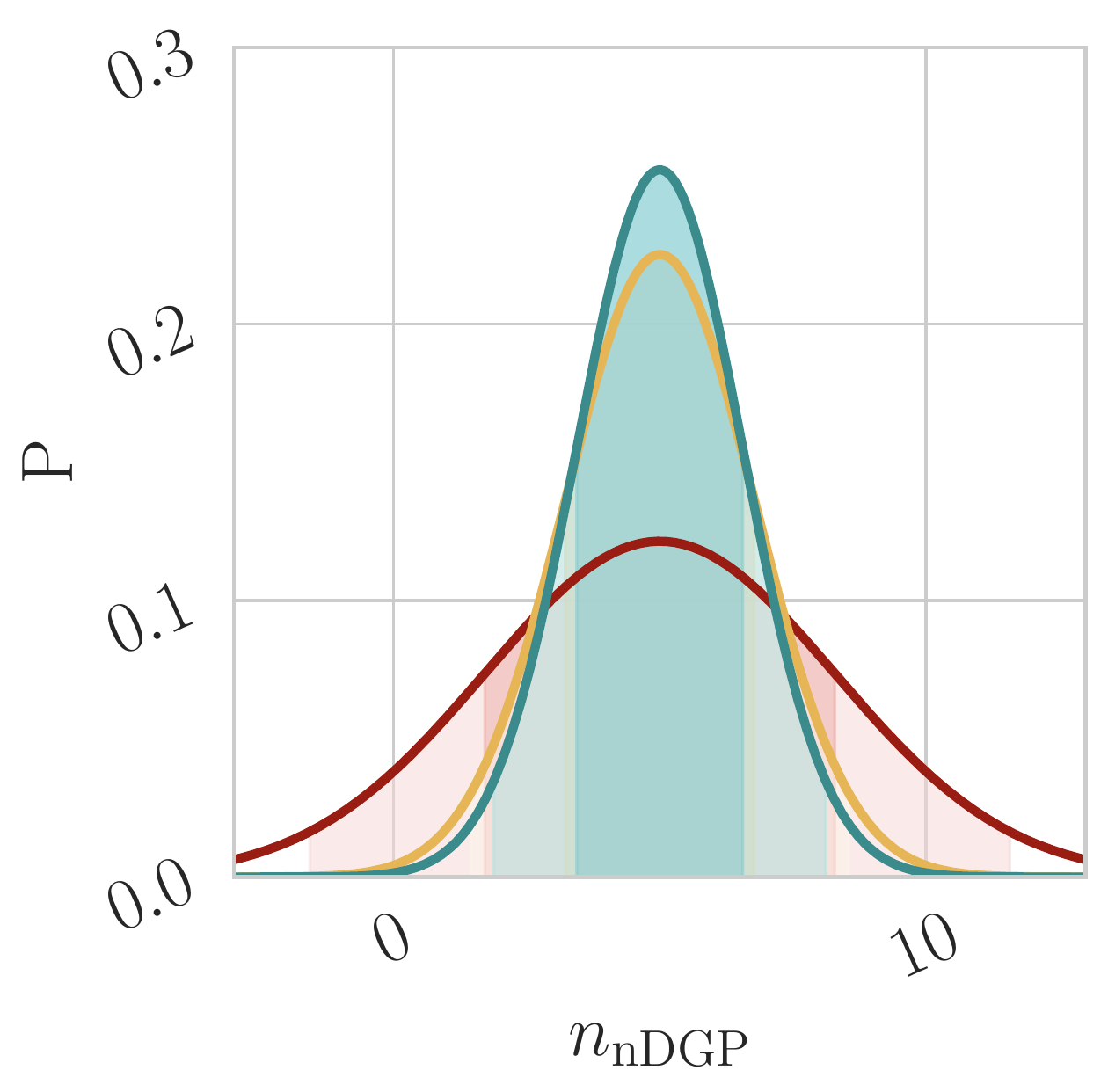}
 }
 \caption{The 1D likelihood distribution for  [Left, (a)] the  near-GR $f(R)$ model, with fiducial value $f_{R0} = 0$ (the value of $n$ becomes redundant) and for the nDGP model with fiducial values [Center, (b)] $n_\text{nDGP} = 1$ and [Right, (c)] $n_\text{nDGP} = 5$, under a Gaussian assumption in the Fisher forecast. The constraints from [red] galaxy clustering (ELG+LRG), [yellow] cluster abundances and [cyan] the two combined are shown, with the darker fill-in shades denoting the corresponding $1$-$\sigma$ $(68\%)$ confidence levels, lighter shades the $2$-$\sigma$ $(95\%)$ confidence levels.}
 \label{fig:constraints_combined}
 \end{figure*}

We now present our findings in figures. We first examine the near-GR $f(R)$ case in Fig.~\ref{fig:constraints_combined} (a), which is conveniently placed alongside the corresponding nDGP constraints, due to that they are all one-dimensional. As we mentioned, the relative impacts of the cluster abundance and galaxy clustering constraints on $f_{R0}$ are comparable for each dataset, and the combination provides a $\sim 25\%$ improvement in the $1$-$\sigma$ constraint relative to that from cluster abundances alone. This result also implies that a $2$-$\sigma$ ($ 95\%$ confidence) upper limit of $f_{R0} \leq5.68\times 10^{-7}$ can be placed for $f_{R0}$ when we take the fiducial scenario as GR ($f_{R0}=0$). 

\begin{figure*}
\subfloat[$f(R)$: F5  (fiducial $\text{log}_{10}(f_{R0}) = -5$)]{%
  \includegraphics[scale = 0.36]{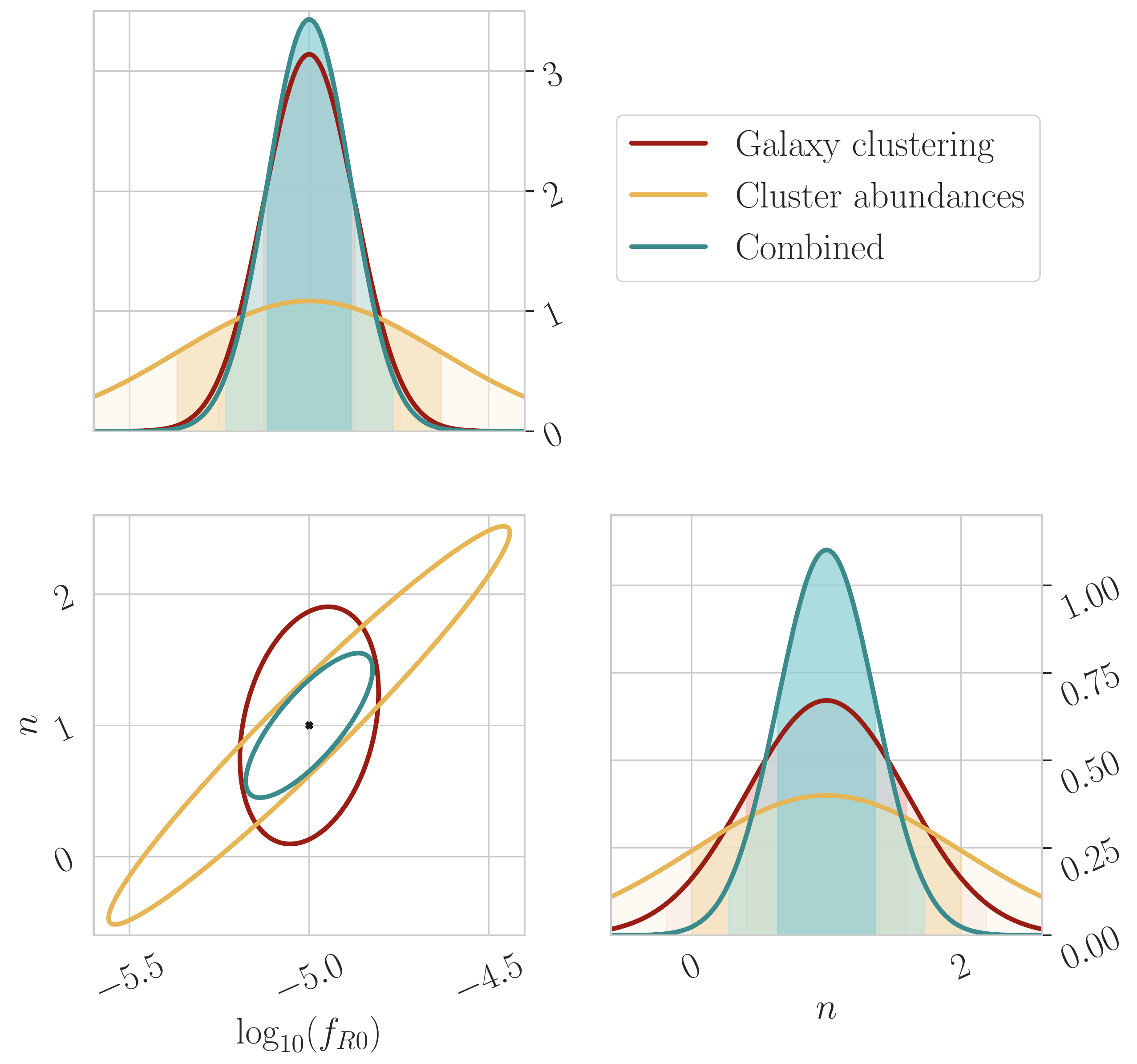}%
}
\subfloat[$f(R)$: F6 (fiducial $\text{log}_{10}(f_{R0}) = -6$)]{%
  \includegraphics[scale = 0.36]{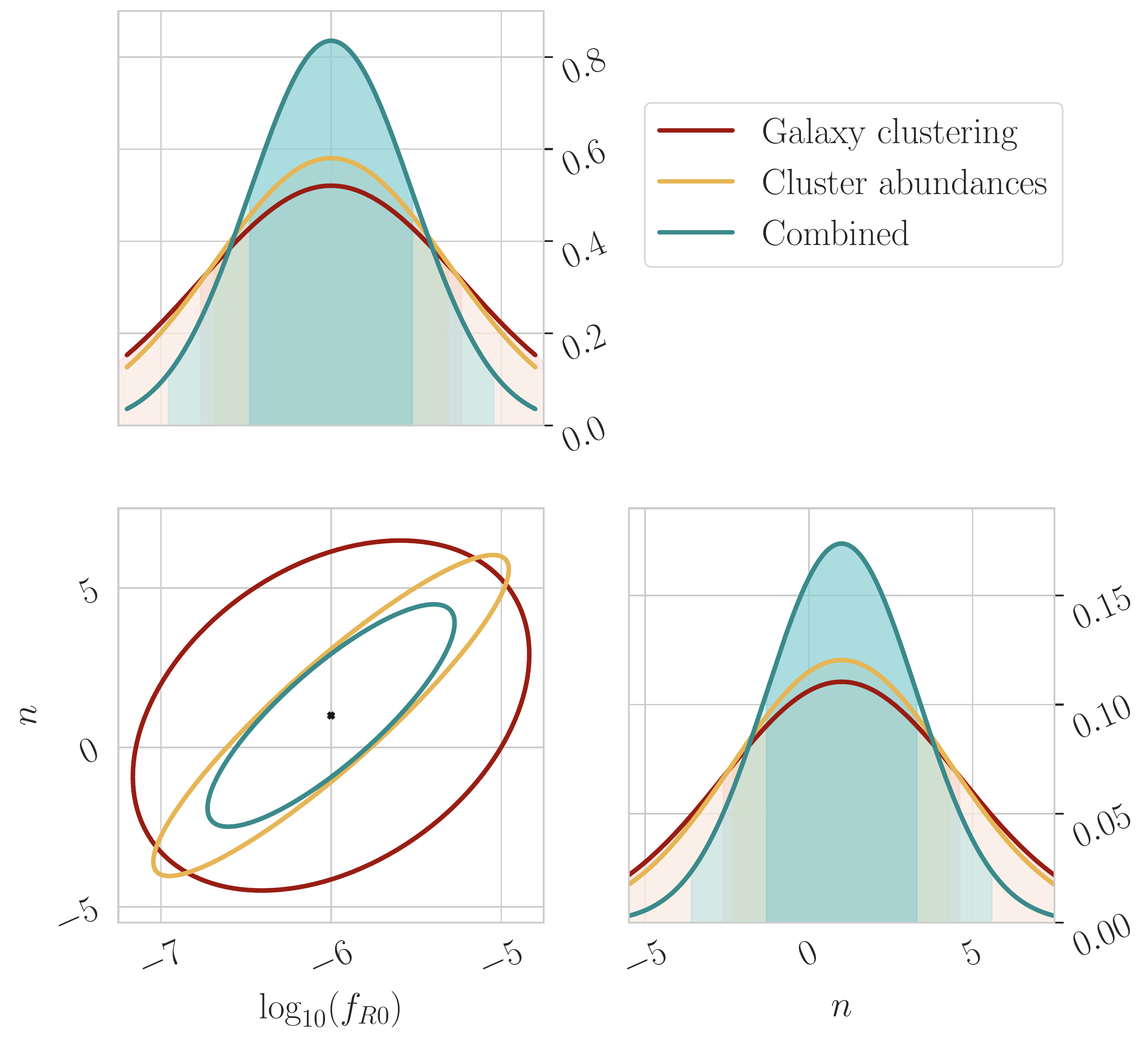}%
}
\caption{The constraints on the [Left] F5 and [Right] F6 $f(R)$ scenarios are shown from [red] galaxy clustering, [yellow] cluster abundances and [cyan] the two combined. The covariance ellipses in the $\{\text{log}_{10}(f_{R0}), n\}$ parameter space indicating the joint $1$-$\sigma$ $(68\%)$ confidence levels, and their respective marginalized 1D Gaussian likelihoods are shown for each scenario, with the darker fill-in shades denoting the corresponding $1$-$\sigma$ $(68\%)$ confidence levels, lighter shades the $2$-$\sigma$ $(95\%)$ confidence levels.}
\label{fig:constraint_plots_fR}
\end{figure*}

We then present the 2D $\{\text{log}_{10}(f_{R0}),n\}$ constraints for the F5 and F6 $f(R)$ scenarios for the two datasets in Fig. \ref{fig:constraint_plots_fR}, with the corresponding combined 1D constraints summarized in Table~\ref{tab:single_param_constraints_MG}. The choice of our prior space and the corresponding caveats have been discussed in \ref{sec:Fisher_method}, and also noted in the captions. For F5 with cluster abundances we find that there is a strong degeneracy in the $\text{log}_{10}(f_{R0})$-$n$ plane but with a well-measured combination in the direction orthogonal to the degeneracy. This phenomenon has been tested to be relatively stable across the higher and lower redshift ranges. In contrast, constraints from galaxy clustering do not show significant degeneracies in this parameter space and provide tighter constraints on $\text{log}_{10}(f_{R0})$ and $n$ separately, but with an overall likelihood ellipse that is wider (the best constraint in the 2D plane is weaker than for the cluster abundances). In combination, the galaxy clustering constraints help break the degeneracy from the cluster abundance data, and drive the constraints on $\text{log}_{10}(f_{R0})$. The constraints on $n$ are improved, relative to those from the galaxy clustering, by a factor of $\sim2$. For F6, the constraints from both cluster abundances and galaxy clustering are weaker, but the galaxy clustering constraints on $\text{log}_{10}(f_{R0})$ are comparable to those from the cluster abundances. The net effect of combining the two datasets is less significant than for F5; however, we see improvements of $\sim30\%$ in constraints on both $\text{log}_{10}(f_{R0})$ and $n$.

The constraints on the nDGP model parameter are also shown in Fig.~\ref{fig:constraints_combined}. Here we find that the cluster abundances drive the constraints for both fiducial scenarios and the galaxy clustering plays a minimal role in affecting improvements, which possibly is also due to the scale-independence that nDGP features.

Spanning the two most popular classes of screening in the literature, through the representative $f(R)$ and nDGP MG models, our detailed analysis overall serves to highlight the ways in which the upcoming precise observations of redshift-space galaxy clustering and cluster abundances will enable us to probe the landscape of dark energy and MG parametrizations in the next 10 years. The nDGP model realizes the Vainshtein screening mechanism, which is harder to constrain using other astrophysical probes, in comparison to the chameleon screening of $f(R)$. Here we find that the cluster abundances are better able to constrain the scale-independent effects of the nDGP model, while the galaxy clustering provides tighter constraints on the scale-dependent $f(R)$ scenario, and could break degeneracies when combined with cluster abundances. This complementarity of the two techniques in constraining these models, and the potential for cluster abundances to constrain nDGP, are important outcomes from our findings.


\section{Discussion} 
\label{sec:disc}

In this work we performed a detailed study of our ability to constrain the large-scale properties of gravity with a combination of two promising probes of the LSS: galaxy clustering from spectroscopic observations by DESI, as well as cluster abundances from tSZ observations by the Simons Observatory.

For galaxy clustering, we employ the Gaussian Streaming Model with Lagrangian Perturbation Theory (LPT) to predict the anisotropic redshift-space 2-point correlation function of biased tracers, which was recently generalized to support predictions for MG parametrizations. We apply the model to predict the multipoles of the RSD correlation function for the ELG and the LRG DESI spectroscopic galaxy samples, as well as their corresponding covariance matrices. Regarding cluster abundances, we use the amplitude of density fluctuations, $\sigma_8(z)$, obtained by tSZ-selected galaxy clusters, as a window into the nature of the underlying gravity model, expanding upon recent detailed model-independent studies in the context of standard cosmological parameters.

We employ the Fisher forecast formalism to obtain a set of joint constraints on  two  widely-studied MG models, the Hu-Sawicki $f(R)$ and the nDGP gravity models. We demonstrate that the two independent probes complement each other in constraining the Hu-Sawicki $f(R)$ model parameters, for a near-GR fiducial scenario, as well as varying degrees of deviation away from a $\Lambda$CDM background. We find that the tightest constraints are obtained in the large-deviation F5 scenario, at the level of a $\sim 2\%$ forecasted joint constraint on the $\text{log}_{10}(f_{R0})$ parameter, with the ELGs serving as the primary source of discriminating power on the galaxy clustering side. The constraining power of both probes is primarily derived from their corresponding lower redshift snapshots, when the MG deviations are more pronounced overall. We also consider the full 2D parameter space, $\{\text{log}_{10}(f_{R0}),n\}$, for the Hu-Sawicki model, and place corresponding constraints.

In a similar manner, we find that the interplay between the cluster abundance and galaxy clustering observables can be utilized to constrain the parameter space of the nDGP gravity scenario. We forecast a combined relative constraint of $\sim 9\%$ in the $n_\text{nDGP}=1$ case and that the cluster abundance observations would principally drive these constraints. This, and the opposite phenomenon that galaxy clustering drives the constraints for the F5 scenario, are potentially due to the fact that the $f(R)$ model is scale-dependent in contrast to nDGP in linear and quasi-linear scales, which might also explain the relatively balanced constraining powers of the two observables for F6 and near-GR, where the scale-dependence is weaker than in F5. 

There are many possible ways in which one can expand upon this line of inquiry. For galaxy clustering, the accuracy of our model can be further improved by including the 1-loop corrections of LPT \citep{Valogiannis:2019nfz} into the GSM prediction, as well as by introducing effective field theory corrections to account for non-perturbative small-scale physics. Such an approach would also need to be combined with a suitably improved treatment of the clustering covariance matrix, which we assumed to be Gaussian in the present work. Furthermore, it would be very interesting to also explore the constraining power of the Fourier space counterpart of the two-point function, the redshift space power spectrum, obtained either by analytical approaches (see e.g. \citep{Aviles:2020wme}) or through emulators \citep{Ramachandra:2020lue}. For galaxy cluster abundances, extending this work to a full Fisher analysis with MG requires halo mass functions \citep[e.g.,][]{Tinker08} with fitting formulas in their {\it ansatz} for the Hu-Sawicki $f(R)$ and the nDGP gravity models. We expect these constraints to further improve with CMB-S4 cluster abundances in combination with photometric and weak gravitational lensing observations by Stage-IV surveys such as the V. Rubin Observatory LSST \citep{Abell:2009aa,Abate:2012za}. Finally, it would be interesting to use a Markov Chain Monte Carlo forecasting approach to see how non-Gaussianities in the posterior likelihood impact the constraints and degeneracies we present.

In the near future, synergies between new cosmological surveys will allow us to explore the vast landscape of dark energy and MG scenarios, and shed new light on the nature of cosmic acceleration. Our work serves to highlight the great promise held in such considerations, as well as the optimal ways in which the vast amounts of upcoming observations could be utilized. 

\begin{acknowledgments}
The authors would like to thank Mathew Madhavacheril for helpful discussions regarding the cluster abundances analysis, as well as code resources that helped plotting the covariance ellipses. Georgios Valogiannis would like to thank Sukhdeep Singh and Yin Li for useful discussions on analytical models of covariance matrices. This work is not an official Simons Observatory paper.

Georgios Valogiannis' work is supported by NSF grant AST-1813694. Georgios Valogiannis and Rachel Bean acknowledge support from DoE grant DE-SC0011838, NASA ATP grant 80NSSC18K0695, NASA ROSES grant 12-EUCLID12-0004 and funding related to the Roman High Latitude Survey Science Investigation Team. Nicholas Battaglia acknowledges support from NSF grant AST-1910021.

\end{acknowledgments}

\newpage
\appendix
\label{sec:app}
\section*{Appendix}
\section{Covariance matrix calculation} 
\label{App:AppendixA}

This Appendix provides the details of the analytical model we use to evaluate the covariance matrix of the multipoles of the anisotropic correlation function. We begin with the known expression for the Poisson error matrix of the power spectrum, assuming Gaussian density perturbations \citep{Feldman:1993ky,Meiksin:1998mu}:
\begin{equation}
\begin{aligned}\label{Covgen}
&Cov\left[P(\bold{k}),P(\bold{k}')\right] = \\
& \frac{(2\pi)^3}{V_s}\left(P(\bold{k}+\frac{1}{n}\right)^2\left(\delta_D(\bold{k}-\bold{k}') + \delta_D(\bold{k}+\bold{k}') \right)\\
&+ \frac{1}{n^2 V_s}\left[P(|\bold{k}-\bold{k}'|) + P(|\bold{k}+\bold{k}'| + 2 P(\bold{k}) +  2 P(\bold{k}') \right] \\
& + \frac{1}{n^3 V_s},
\end{aligned}
\end{equation}
with $n$ the number density of the galaxies in a given sample, $V_s$ the survey of the volume and $\delta_D$ the Dirac delta-function. The second and third lines on the r.h.s of Eqn.~(\ref{Covgen}) encode the Poisson shot noise contributions to the covariance matrix~\citep{Cohn:2005ex}. Eqn.~(\ref{Covgen}) has neglected contributions from non-linear gravitational evolution \citep{Meiksin:1998mu,Scoccimarro:1999kp,Cooray:2000ry,
Barreira:2017kxd}, super sample covariance \citep{Hamilton:2005dx,
Hu:2002we,
Takada:2013wfa,Li:2017qgh,
Barreira:2017fjz} and effects of the survey nontrivial window function \citep{Li:2018scc}. 

Our goal is to Fourier transform the result~(\ref{Covgen}), so that we obtain the configuration space equivalent expression for the covariance matrix of the anisotropic correlation function. In the simpler case of real space considerations, with the correlation function being isotropic,~\citep{Cohn:2005ex} demonstrated that, by angle-averaging the Fourier transform of~(\ref{Covgen}), the oscillatory Bessel function dependencies can be eliminated (unlike in the RSD case, as shown below), and a more compact expression is possible. ~\citep{Cohn:2005ex} also found the Poisson shot-noise contributions to be diagonal for the correlation function. In redshift space, which is what we are interested in in this work, the equivalent configuration space expression for~(\ref{Covgen}) has been derived in \citep{White:2014naa,Grieb:2015bia}, assuming only Gaussian shot-noise contributions (i.e. neglecting the second and third lines in the r.h.s of~(\ref{Covgen})), and is the following: 
\begin{equation}
\begin{aligned}\label{Covxishort}
& Cov\left[\xi_{l_1}(s_i),\xi_{l_2}(s_j)\right] = \\  & \frac{i^{l_1+l_2}}{2\pi^2}\int_0^{\infty}k^2 \sigma^2_{l_1 l_2}(k)j_{l_1}(k s_i)j_{l_2}(k s_j)dk,
\end{aligned}
\end{equation}
where we defined the multipole per-mode covariance:
\begin{equation}
\begin{aligned}\label{Covpermode}
& \sigma^2_{l_1 l_2}(k) = \frac{(2 l_1+1)(2 l_2 +1)}{V_s}\\  & \times \int_{-1}^{1} \left[P(k,\mu_k) + \frac{1}{n} \right]^2 L_{l_1}(\mu_k)L_{l_2}(\mu_k)d\mu_k,
\end{aligned}
\end{equation}
where $j_{l_1}(k s_i),j_{l_2}(k s_j)$ are the spherical Bessel functions of the first kind. Poisson shot-noise contributions can potentially become significant, as pointed out by~\citep{Cohn:2005ex}. To that end, we proceed to expand the expression (\ref{Covxishort}) to also include the Poisson terms to the shot-noise contributions, just as in the real-space version of~\citep{Cohn:2005ex}. To do so, we first adopt our convention for the Fourier transformation, applied on the correlation function:
\begin{equation}\label{eq:FT}
\xi(\bold{s}) = \int \frac{d^3 k}{(2\pi)^3} e^{i \bold{k}\cdot \bold{s}} P(\bold{k}),
\end{equation}
and label the terms reflecting the Poisson shot-noise contributions in~(\ref{Covgen}) (second and third lines of r.h.s) as $Cov\left[P(\bold{k}),P(\bold{k}')\right]\bigg\rvert_{\text{Poisson}}$. Fourier transforming both sides then gives\footnote{The Fourier transformation of the r.h.s gives rise to additional terms involving delta-functions, as in~\citep{Cohn:2005ex}, that only contribute at separations $r=0$, and are thus dropped.}
\begin{equation}\label{eq:FTinter}
\begin{aligned}
& Cov\left[\xi(\bold{s}_i),\xi(\bold{s}_j)\right]\bigg\rvert_{\text{Poisson}} = \\ & \frac{2}{n^2 V_s}\xi(\bold{s}_i) \delta_D(\bold{s}_i-\bold{s}_j).
\end{aligned}
\end{equation}
Finally, we aim to project out the correlation function multipoles, for which we integrate the $\xi$ terms on the l.h.s above (after multiplying both sides with the appropriate Legendre polynomials), as in (\ref{mult}), which gives
\begin{equation}\label{eq:FTfin}
\begin{aligned}
& Cov\left[\xi_{l_1}(s_i),\xi_{l_2}(s_j)\right]\bigg\rvert_{\text{Poisson}} = \\ & \frac{(2 l_1 +1)(2 l_2 +1)}{n^2 V_s 4 \pi s_i^2} \delta_D(s_i-s_j)\int_{-1}^{1}\xi(s_i,\mu_s) L_{l_1}(\mu_s)L_{l_2}(\mu_s) d \mu_s,
\end{aligned}
\end{equation}
where we have made use of the delta-function property:
\begin{equation}
\delta_D(\bold{s}_i-\bold{s}_j) = \frac{\delta_D(s_i-s_j)}{s_i^2}\delta_D(\Omega_{i}-\Omega_j),
\end{equation}
with $\Omega$ denoting the corresponding solid angles. Combining (\ref{eq:FTfin}) with (\ref{Covxishort}), we arrive at the desired result:
\begin{equation}
\begin{aligned}\label{Covxifinal}
& Cov\left[\xi_{l_1}(s_i),\xi_{l_2}(s_j)\right] = \\  & \frac{i^{l_1+l_2}}{2\pi^2}\int_0^{\infty}k^2 \sigma^2_{l_1 l_2}(k)j_{l_1}(k s_i)j_{l_2}(k s_j)dk + \\
& \frac{(2 l_1 +1)(2 l_2 +1)}{n^2 V_s 4 \pi s_i^2} \delta_D(s_i-s_j)\int_{-1}^{1}\xi(s_i,\mu_s) L_{l_1}(\mu_s)L_{l_2}(\mu_s) d \mu_s,
\end{aligned}
\end{equation}
which is the equation we use to evaluate the covariance matrix of the multipoles of $\xi$ in this work. The last term in Eqn. (\ref{Covxifinal}) expands the Gaussian expression (\ref{Covxishort}) of \citep{White:2014naa,Grieb:2015bia} in order to also capture the Poisson shot-noise contributions in the anisotropic case, and exhibits the same diagonal nature as the corresponding real space expression of Eqn. (32) in ~\citep{Cohn:2005ex}, which it recovers in the limit of isotropy. The shot-noise terms in (\ref{Covxifinal}) are further divided by the bin windth, $\Delta s$, in order to avoid overestimating the error predictions, as in \citep{Cohn:2005ex,PhysRevD.77.043525}. 

To summarize our procedure, after getting an analytical prediction for the RSD correlation function for our desired cosmology from Eqn. (\ref{xiGSM}), we use it to predict the covariance matrix from Eqn.~(\ref{Covxifinal}) (combined with the input from (\ref{Covpermode})). An intermediate step is to Fourier Transform to also obtain $P(k,\mu_k)$ from $\xi(s,\mu_s)$, which is required in Eqn. (\ref{Covpermode}), and can be easily performed with the publicly available package $\texttt{mcfit}$\footnote{\url{https://github.com/eelregit/mcfit}}. The integrals involving spherical Bessel functions in (\ref{Covxishort}) can also be conveniently performed by utilizing the same package.

\newpage
%


\end{document}